\documentstyle[12pt]{article}

\newcommand{\sect}[1]{\setcounter{equation}{0}\section{#1}\indent}

\newcommand{\EQ}{\begin{equation}}
\newcommand{\EN}{\end{equation}}
\newcommand{\bea}{\begin{eqnarray}}
\newcommand{\ena}{\end{eqnarray}}
\newcommand{\vs}[1]{\vspace{#1 mm}}

\renewcommand{\a}{\alpha}

\renewcommand{\d}{\delta}
\newcommand{\e}{\epsilon}

\newcommand{\shalf}{\frac{1}{2}}
\newcommand{\pa}{\partial}
\newcommand{\tri}{{\small $\triangle$}}
\newcommand{\dz}{\frac{dz}{2\pi i}}
\newcommand{\ra}{\rangle}
\newcommand{\lan}{\langle}
\newcommand{\D}{{\cal D}}
\newcommand{\uda}{\nearrow \kern-1em \searrow}

\begin{document}

\topmargin 0pt
\oddsidemargin 5mm

\renewcommand{\Im}{{\rm Im}\,}
\newcommand{\NP}[1]{Nucl.\ Phys.\ {\bf #1}}
\newcommand{\PL}[1]{Phys.\ Lett.\ {\bf #1}}
\newcommand{\CMP}[1]{Comm.\ Math.\ Phys.\ {\bf #1}}
\newcommand{\PR}[1]{Phys.\ Rev.\ {\bf #1}}
\newcommand{\PRL}[1]{Phys.\ Rev.\ Lett.\ {\bf #1}}
\newcommand{\PTP}[1]{Prog.\ Theor.\ Phys.\ {\bf #1}}
\newcommand{\PTPS}[1]{Prog.\ Theor.\ Phys.\ Suppl.\ {\bf #1}}
\newcommand{\MPL}[1]{Mod.\ Phys.\ Lett.\ {\bf #1}}
\newcommand{\IJMP}[1]{Int.\ Jour.\ Mod.\ Phys.\ {\bf #1}}

\begin{titlepage}
\setcounter{page}{0}
\begin{flushright}
OS-GE 28-93\\
OU-HET 173\\
hep-th/9305179
\end{flushright}

\vs{5}
\begin{center}
{\Large BRST ANALYSIS OF PHYSICAL STATES IN TWO-DIMENSIONAL BLACK HOLE}
\vs{10}

{\large Katsumi Itoh,$^{1}$ Hiroshi Kunitomo,$^2$ Nobuyoshi Ohta$^{3}$}\\
and \\
{\large Makoto Sakaguchi$^2$}\\
\vs{5}
$^1${\em Faculty of Education, Niigata University, Niigata 950-21, Japan\\
and Institute for Theoretical Physics, University of California \\
Santa Barbara, CA 93106, U. S. A.}\\
\vs{5}
$^2${\em Department of Physics, Osaka University, Toyonaka, Osaka 560,
Japan}\\
\vs{5}
$^3${\em Institute of Physics, College of General Education,
Osaka University \\ Toyonaka, Osaka 560, Japan} \\
\end{center}
\vs{10}

\centerline{{\bf{Abstract}}}

We study the BRST cohomology for $SL(2,R)/U(1)$ coset model, which
describes an exact string black hole solution. It is shown that the
physical spectrum could contain not only the extra discrete states
corresponding to those in $c=1$ two-dimensional gravity but also many
additional new states with ghost number $N_{FP}= -1 \sim 2$. We also
discuss characters for nonunitary reps. and the relation
of our results to other approaches.

\end{titlepage}
\newpage
\renewcommand{\thefootnote}{\arabic{footnote}}
\setcounter{footnote}{0}

\sect{Introduction}
\indent
Two-dimensional gravity or string theory has been widely studied
as the only viable candidate for consistent theory of quantum gravity
coupled to matter. In this approach, the study of curved space-time as
string backgrounds should be useful to investigate
how gravity behaves at the Planck scale.

Particularly relevant in this analysis is the background of black holes.
Indeed, many of the most interesting and important questions in
quantum gravity concern the physics of black holes. Some insight into
these problems has already been obtained from two-dimensional models of
black holes~\cite{WA,BLA}. However, most of the works employed
semi-classical approximations and have been unable to give a consistent
picture of quantum theory. Clearly it is necessary to understand full
quantum effects in order to get such a picture.

Recently there has been substantial progress in developing
nonperturbative techniques to handle strings in nontrivial
backgrounds~\cite{BDG}, and this has stimulated the remarkable advance
in our understanding of the continuum theory of two-dimensional gravity.
In particular, it has been found that there are an infinite number of
extra states at discrete values of momenta~\cite{LIZ,MUK} which
represent the $W_\infty$ symmetry of the theory~\cite{WIT1,KP,WIN,RIN}.
It is expected that these states play very important roles in our
understanding of the structure of the theory.

String backgrounds may be most effectively studied with conformal field
theory.  However most of the existing works have been restricted to
compact curved space. In order to study string theory in general curved
backgrounds, it will be useful to examine noncompact coset
models~\cite{BN,DLP,HWA}. In particular, it has been shown that a gauged
$SL(2,R)$ Wess-Zumino-Witten model is a conformal field theory for
two-dimensional string in a black hole background~\cite{WIT2,BK}. It is
thus quite important to understand the structure of this coset model.
To this aim, it is first necessary to clarify the physical spectrum in
the theory. This problem has been examined by several groups by
adopting the BRST approach~\cite{DN,EY,CL}. These analyses indicate
that there are also an infinite number of discrete states in the theory.
However, it seems that the precise structure of the physical spectrum
in $SL(2,R)/U(1)$ has not been clarified yet.

The main purpose of this paper is to compute the BRST cohomology and
derive the spectrum for the $SL(2,R)/U(1)$ coset model. We study the
BRST cohomology in two steps. We first compute the cohomology of
the $SL(2,R)/U(1)$ parafermion (PF) modules with irreducible base reps.
The results agree with those obtained by Distler and Nelson~\cite{DN}.
These PF modules, however, are in general reducible and contains null
states. To get irreps. of the coset, we must subtract submodules
generated from the null states. We next identify the null states and
examine how the cohomology is changed if one takes them into account.
As a result, we find that there are more discrete physical states than
those given in ref.~\cite{DN}. We also derive the character formulae
for these reps. based on our knowledge of the null states.

It would be appropriate to explain why we find more discrete states
here than $c=1$ matter coupled to gravity. The situation is best
understood in the free boson realization. After the division by
$U(1)$, our coset theory may be described by two bosons, which
look quite similar to the $c=1$ matter and Liouville fields.
Thus, if we study the spectrum in Fock spaces for these bosons
and the ghosts, we are bound to get the same spectrum as $c=1$
matter coupled to gravity.  All the other states are in a BRST doublet,
$Q_B \psi = \chi$. In our discussion, we divide Fock spaces first to
the PF modules with irreducible base reps. and next to PF irreps.
In the first step there are cases that $\psi$ and $\chi$ are in
different modules on different irreducible base reps. Thus each of
the doublet becomes a physical state in each module. This gives us
new discrete states obtained in sect.~3.  In the second step $\chi$
$(\psi)$ may be null states and this makes $\psi$ $(\chi)$ physical.
The real physical states for string in the black hole background must
be the subset of those in sect.~5.  There is also a possibility to
find more than one string theory with the same black hole geometry.

In sect.~2, we discuss $SL(2,R)$ algebra, the coset $SL(2,R)/U(1)$ and
their free field realization~\cite{WZ}. In this paper, we will consider
the Euclidean black hole: namely, we divide out the negative metric
field in the coset construction. Another formulation to gauge away the
$U(1)$ with the BRST formalism is explained in appendix A. In sect.~3,
we compute the BRST cohomology for the $SL(2,R)/U(1)$ PF Verma modules
(this is the case considered in~\cite{DN}). In computing the cohomology
of the coset, we use the cohomology analysis for $2D$ gravity~\cite{LIZ}.
A brief discussion of exact sequences necessary to our analysis is
relegated to appendix B. In sect.~4, we construct null states in the
$SL(2,R)$ algebra and $SL(2,R)/U(1)$ coset by using the free field
realization. The result is consistent with the determinant formula
of Kac and Kazhdan~\cite{DLP,KK}. This result is then used in sect.~5
to derive the physical spectrum for the $SL(2,R)/U(1)$ coset model.
Sect.~6 is devoted to discussions of the character formulae
for various reps.~\cite{CHA}. The relation of our results
to other approaches~\cite{EY} is also illustrated by taking simple
examples of the discrete physical states.

\sect{$SL(2,R)$ and free field realization}
\indent
In this section, we briefly summarize the reps. and free field
realizations for $SL(2,R)$ current and PF algebras.

The $SL(2,R)$ algebra has three generators with
operator product expansions (OPEs)
\EQ
J^i(z)J^j(w) \sim  \frac{Kg^{ij}/2}{(z-w)^2}+
\frac{i{\e^{ij}}_k J^k(w)}{z-w},
\EN
where the metric $g_{ij}=$ diag. $(1,1,-1)$ and
$K$ is the level of this algebra. If we set
\EQ
J^\pm(z)=J^1\pm iJ^2,
\EN
the OPEs take the form
\bea
J^+(z)J^-(w) &\sim& \frac{K}{(z-w)^2}-\frac{2}{z-w} J^3(w),\nonumber \\
J^3(z)J^\pm(w) &\sim& \frac{\pm}{z-w} J^\pm(w), \\ \nonumber
J^3(z)J^3(w) &\sim& \frac{-K/2}{(z-w)^2},
\ena
and we have the hermiticity
\bea
(J^+(z))^\dagger &=& J^- \left( \frac{1}{\bar z}\right)(\bar z)^{-2},
 \nonumber\\
(J^3(z))^\dagger &=& J^3 \left( \frac{1}{\bar z}\right)(\bar z)^{-2}.
\ena
The stress tensor is given by
\EQ
T^{SL(2)}(z) = \frac{1}{2(K-2)} :[J^+(z)J^-(z)+J^-(z)J^+(z)-
 2 (J^3(z))^2]:,
\EN
which satisfies the conformal OPE with central charge
\EQ
c=\frac{3K}{K-2}.
\EN

In order to construct reps. of the $SL(2,R)$ algebra,
it is useful to introduce the mode operator $J^i_n$ by
\EQ
J^i(z)=\sum_nJ^i_nz^{-n-1}.
\EN
The hermiticity (2.4) implies
\EQ
(J^+_n)^\dagger = J^-_{-n},\qquad (J^3_n)^\dagger = J^3_{-n}.
\EN

By the standard procedure, we find the commutation
relations from the OPEs (2.3)
\bea
[J^+_n, J^-_m] &=& Kn \d_{n+m,0} -2J^3_{n+m}, \nonumber \\
{}[J^3_n, J^\pm_m] &=& \pm J^\pm_{n+m}, \\ \nonumber
[J^3_n, J^3_m] &=& -\frac{K}{2}n \d_{n+m,0},
\ena
as well as the Virasoro operator
\bea
L^{SL(2)}_0&=&\frac{1}{K-2}\left[ {\bf J}^2
+\sum^\infty_{n=1}(J^+_{-n}J^-_n+J^-_{-n}J^+_n-2J^3_{-n}J^3_n)
\right], \nonumber \\
{\bf J}^2 &\equiv& \frac{1}{2}(J^+_0J^-_0+J^-_0J^+_0)-(J^3_0)^2.
\ena

The reps. of the $SL(2,R)$ current algebra are
built over irreps. of the global
$SL(2,R)$ (generated by $J^\pm_0$ and $J^3_0$)
by applying the $J^i_{-n}, n>0$, in all possible ways.

Since we consider a target space with the time direction, we do not
restrict ourselves to unitary reps. of the $SL(2,R)/U(1)$ coset.
Instead, we take {\it hermitian reps.}, where the
currents are realized as hermitian operators and the inner product is
non-degenerate~\cite{DN}.  The unitarity should be discussed in the
physical subspace as in the critical string theory.

The reps. of the global $SL(2,R)$ are constructed in exactly the same
way as those of $SU(2)$. These are characterized by spin
${\bf J}^2=-j(j-1)$. It follows that the reps. with $j$ and $1-j$ are
equivalent. To avoid the double counting, we restrict $j\ge 1/2$.

The $SL(2,R)$ reps. are obtained by determining the eigenvalues
of $J^3_0$. We first consider a state $|m\ra$ with $J^3_0=m$ and then
act on this state by the ladder operators $J^\pm_0$. After some
algebra, one finds that the hermitian irreps.
of the global $SL(2,R)$ fall into the following four classes:

\begin{enumerate}
\item Lowest weight (LW) reps.:
${\tilde \Delta}^+\equiv \{|j\ra,|j+1\ra,|j+2\ra, \cdots \}$ or
$\Delta^+\equiv \{|-j+1\ra, \\ |-j+2\ra,|-j+3\ra, \cdots \}$,
$j \ge 1/2$ with $J^-_0|j\ra=0$ or $J^-_0|-j+1\ra=0$;
\item Highest weight (HW) reps.:
${\tilde \Delta}^-\equiv\{|-j\ra,|-j-1\ra,|-j-2\ra,
\cdots \}$ or $\Delta^-\equiv\{|j-1\ra,\\ |j-2\ra,|j-3\ra,\cdots \}$,
$j\ge 1/2$ with $J^+_0|-j\ra=0$ or $J^+_0|j-1\ra=0$;
\item Double-sided reps.: $U\equiv\{
|-j+1\ra,|-j+2\ra, \cdots, |j-2\ra,|j-1\ra\}$, $j= 1/2, 1, 3/2, \cdots,$
with\footnote{For $j=\shalf$, $J_0^\pm$ are interchanged in these
conditions.} $J^-_0|-j+1\ra=0$, $J^+_0|j-1\ra=0$;
\item Continuous reps.: $C\equiv\{|k+\phi_0
\ra\}$ with integer $k$ and $1>\phi_0\ge 0$ and $j\ge 1/2$,\footnote{
For the continuous reps., $j$ is allowed to take the complex
value $j=1/2+i\lambda$ with $\lambda>0$. In this paper this
possibility will not be studied
since it cannot satisfy the on-shell condition.}
\end{enumerate}
where we indicate only $J^3_0$ eigenvalues. These reps. provide the
bases to build up reps. of the $SL(2,R)$ current algebra and will be
called base reps. in what follows.

In order to make a connection with the two-dimensional gravity coupled
to matter, it is useful to bosonize the currents as \cite{WZ}
\bea
J^\pm &=& \left(i\sqrt{\frac{K}{2}}\pa\phi^M \mp
\sqrt{\frac{K-2}{2}}\pa\phi^L\right) \exp
\left( \pm\sqrt{\frac{2}{K}}(i\phi^M+\phi^3) \right), \nonumber \\
J^3 &=& -\sqrt{\frac{K}{2}}\pa\phi^3,
\ena
where $\phi^i(z)\phi^j(w) \sim -\d^{ij}\ln(z-w)$. It is easy to see
that the OPEs in (2.3) are satisfied by the currents in (2.11).

The base reps. may be constructed in terms of vertex operators,
\EQ
U_{jm}(z)= \exp\left[ \sqrt{\frac{2}{K-2}}j\phi^L+\sqrt{\frac{2}{K}}m
(i\phi^M+\phi^3) \right],
\EN
as
\EQ
|j,m\ra = U_{j,m}(0)|0\ra,
\EN
which has the dimension $-j(j-1)/(K-2)$ and $J^3_0=m$.
The LW rep.
${\tilde\Delta}^+$ corresponds to the set of values
$\{m=j,j+1,\cdots \}$, while the HW ${\tilde\Delta}^-$ to
$\{m=-j,-j-1, \cdots \}$ with $j$ replaced by $-j$.
Indeed, we find the OPEs
\bea
J^3(z)U_{jm}(w)\sim\frac{m}{z-w}U_{jm}(w), \nonumber \\
J^\pm(z)U_{jm}(w)\sim\frac{m\pm j}{z-w}U_{j,m\pm1}(w),
\ena
which ensure that the reps. split precisely at $m=\pm j$.

A comment is in order. In the free boson realization, it would be
natural to start with a generally reducible rep. for global $SL(2,R)$
rep., $F\equiv\{|k+m_0\ra\}$ with integer $k$ and some $m_0$, and
decompose it into irreps. For example, when $m_0=j$, one sees from eq.
(2.14) that there is a LW state $|j\ra$ in $F$. We thus get a LW rep.
$\tilde{\Delta}^+$ generated from $|j\ra$ by acting $J^+_0$.
On the other hand, we obtain a HW rep. $\Delta^-$ by setting the above
LW rep. equal to zero in $F$. Namely, $\Delta^-$ is defined as the coset
\EQ
\Delta^-=F/{\tilde\Delta}^+.
\EN
This shows that the LW state $|j\ra$ can be interpreted as a null state
in $F$ when we consider the HW rep. $\Delta^-$. In similar manners, one
finds all the irreps. from $F$ depending on whether
$m\pm j$ is integer or not.

The $SL(2,R)/U(1)$ coset model is obtained
by requiring $J^3 \sim 0$. This eliminates the negative
metric field and gives a model corresponding to a Euclidean black
hole~\cite{WIT2}. The stress tensor for this coset is given by
\bea
T(z)&=&T^{SL(2)}(z)+\frac{1}{K}(J^3(z))^2, \nonumber\\
&=&-\frac{1}{2}(\pa\phi^M)^2-\frac{1}{2}(\pa\phi^L)^2
+\frac{1}{\sqrt{2(K-2)}}\pa^2\phi^L.
\ena
The central charge of the coset model is given by
\EQ
c=\frac{3K}{K-2}-1.
\EN
An alternative formulation to give the coset model is discussed in
appendix A.

This coset may be regarded as the standard parafermionic model
defined with PF fields $\psi^\pm(z)$,
\EQ
J^\pm(z)= \sqrt{K} \psi^\pm (z) \exp \left( \pm\sqrt{\frac{2}{K}}
\phi^3(z) \right).
\EN
The base for PF modules is also given by (2.13),
which has the dimension
\bea
h_{jm} &=& -\frac{j(j-1)}{K-2}+\frac{1}{K}m^2 \nonumber\\
 &=& h_{j,-m} = h_{1-j,m}.
\ena
The PF modules may be obtained by restricting $SL(2,R)$ modules
by the conditions, $J^3_n =0, \; n \geq 1$.
We will denote , following ref.~\cite{DN},
the PF reps.  generated on the lowest (highest)
reps. ${\tilde\Delta^+}({\tilde\Delta^-})$ by ${\tilde\D}^+
({\tilde\D}^-)$ and those on the complementary reps.
$\Delta^-(\Delta^+)$ by $\D^-(\D^+)$. The reps.
generated on double-sided $U$ and continuous reps. $C$ are
denoted by ${\cal U}$ and ${\cal C}$, respectively.

In our discussions of the physical spectrum, an important role is
played by the nilpotent operators
\bea
S^\pm &=& \oint\dz \exp \left(  \sqrt{\frac{K-2}{2}}\phi^L
 \pm i\sqrt{\frac{K}{2}} \phi^M  \right),\nonumber\\
(S^+)^2&=&(S^-)^2=0,
\ena
which commute with the currents and hence satisfy
\EQ
[S^\pm, \psi^\pm(z)] = [S^\pm, \psi^\mp(z)] = 0.
\EN
It follows that these operators provide an isomorphism of the PF
modules \cite{DN}.

Similarly to the discussion after (2.14) for global $SL(2,R)$ rep.,
PF Verma modules can be obtained from free boson Fock modules.
We define the vector spaces ${\cal F}^{\pm}_j$ as
\EQ
{\cal F}^\pm_j\equiv \bigoplus_{m=\pm j+{\bf Z}}{\cal F}_{jm},
\EN
with ${\cal F}_{jm}$ being the $\phi^L$ and $\phi^M$ Fock modules over
$U_{jm}$. The PF modules ${\tilde\D}_j^\pm$ and
$\D_j^\pm$ may be defined by
\EQ
{\tilde\D}_j^\pm={\cal F}^\pm_j\cap Ker S^\pm, \;\;\;\;
\D_j^\pm={\cal F}^\mp_j/{\tilde\D}_j^\mp,
\EN
where we have used (2.21) and the fact that the base rep.
$\tilde{\Delta}^\pm$ is characterized as $Ker S^\pm$.
The PF modules $\D_j^{\pm}$ are obtained from vector spaces
${\cal F}^\mp_j$ by setting the states in the complementary modules
${\tilde\D}_j^{\mp}$ equal to zero.

In order to consider dual spaces to these modules ${\cal F}^\pm_j$,
${\tilde\D}_j^\pm$ and $\D_j^\pm$, we note that $\phi^L$ has a
background charge in eq. (2.16).
This leads to the inner product in base reps. as
\EQ
\lan j',m'|j,m\ra =\delta_{j',1-j}\delta_{m',m},
\EN
which implies that the dual space to ${\cal F}_{jm}$ is isomorphic to
${\cal F}_{1-j, m}$:
\EQ
({\cal F}_{jm})^*\cong {\cal F}_{1-j, m}.
\EN
By means of (2.25) and the hermiticity of $\psi^{\pm}$ implied by (2.4)
and (2.18), we can show that the following isomorphisms hold for
the dual spaces:
\bea
({\cal F}^\pm_j)^*&\cong& {\cal F}^\mp_{1-j},\nonumber \\
({\tilde\D}_j^\pm)^*&\cong& \D_{1-j}^\pm,
\ena
where we have used the hermiticity relation $(S^+)^\dagger=S^-$ which
follows from (2.20).

The total Fock space consists of the tensor product of ${\cal F}_{j,m}$
and the ghost Fock space. We introduce the BRST operator
\EQ
Q_B=\oint\dz(cT+bc\pa c).
\EN
Requiring that the BRST charge be nilpotent or the total central charge
be zero, we get
\EQ
K=\frac{9}{4}.
\EN
In the rest of this paper, we restrict our discussions to this value
of $K$ except in sect. 4. Our physical state condition is then
\EQ
Q_B|{\rm phys}\ra=0.
\EN
We decompose the BRST charge as usual with respect to ghost zero modes
\EQ
Q_B = c_0 L_0 + b_0 M + d.
\EN
We note that any physical states are BRST-exact unless they satisfy the
on-shell condition $L_0 = 0$. It is convenient to reduce the physical
subspace by restricting to states annihilated by $b_0$ (known as
relative cohomology). Our physical state condition is then reduced to
\EQ
L_0|{\rm phys}\ra=b_0|{\rm phys}\ra=d|{\rm phys}\ra=0.
\EN
We note that $d^2=0$ in this subspace.

\sect{BRST cohomology over PF Verma modules}
\indent
As the first step to derive the complete relative cohomology for the
irreducible $SL(2,R)/U(1)$  PF reps., let us discuss
the cohomology over PF Verma modules.

By using the parameters $k$ and $l$ defined by
\EQ
j=\frac{1}{4}(k+l+2), \;\; m=\frac{3}{4}(k-l),
\EN
the on-shell condition
\EQ
h_{j,m} + N \equiv -4j(j-1)+\frac{4}{9}m^2+N = 1,
\EN
is cast into
\EQ
N=kl.
\EN
This implies that the on-shell condition is satisfied only if
$k,l\ge 0$ or $k,l\le0$. Here we will study the case $k,l\geq 0$, which
corresponds to the restriction $j\geq 1/2$ as mentioned in sect. 2.
The region $k,l\leq 0$ is obtained by the replacement $(j,m)\to (1-j,m)$
and is isomorphic to dual spaces as explained in sect. 2.

For $N=0$, we find from the known cohomology $H^1({\cal F})={\bf C}$
\cite{LIZ} that we have a physical state called tachyon with
continuous spectrum:
\EQ
H^1({\cal C})={\bf C}, \qquad{\rm for}\;\; k=0 \;\;{\rm or}\;\; l=0,
\;\; (m=\pm (3j-\frac{3}{2})).
\EN
This tachyon state is in the continuous rep. ${\cal C}$ for a generic
momentum. For special momenta such that $j\pm m\in {\bf Z}$, however,
this is in other reps. as discussed below.

For $N>0$, we will show that there are discrete physical states. For
this purpose we use the mathematical tools of exact sequences
which are recapitulated in appendix B. First note that we have the
short exact sequence
\EQ
0 \to {\tilde\D}^\pm_{jm} \to {\cal F}_{jm}
\to \D^\mp_{jm}\to 0.
\EN
Here the first map is the embedding of ${\tilde\D_{jm}}^\pm={\cal F}
^\pm_{jm}\cap Ker S^\pm$ [see (2.23)] into the free boson module.

In the following study of cohomology, we use isomorphism among PF
modules provided by $S^\pm$. $S^\pm$ cannot act on all states with
arbitrary $j$ and $m$. Indeed, the action of $S^+$ is well defined
only for states with
\EQ
m-j = {\rm integer \;\;\; or  } \;\; l=\shalf (k-1)+({\rm integer}),
\EN
and it maps a state at $(k,l)$ to that at $(k+1, l-\shalf)$. Similarly
$S^-$ is well defined for
\EQ
m+j = {\rm integer \;\;\; or} \;\; l= 2k -({\rm odd \; integer}),
\EN
and it changes $(k,l)$ to $(k-\shalf, l+1)$.
On these lines (3.6,7), we are interested in points $(k,l)$, $(k+1,
l-\shalf)$ and $(k-\shalf,l+1)$ with $k, l\in {\bf Z}$. The points
$(k+1,l-\shalf)$ and $(k-\shalf,l+1)$ are related by the maps $S^\pm$
to points $(k,l)$, on which the cohomology of free boson modules is
nontrivial \cite{LIZ}. Consequently the cohomology for PF modules can
be nontrivial on these points. We depict these points and the
directions of the actions of $S^\pm$ in Fig. 1.

{}From the definition (2.23) of PF modules and the nilpotency of $S^\pm$,
it is easy to see that maps $S^\pm$ give an isomorphism among PF modules
\EQ
\D^\mp_{j,m}\cong
{\tilde\D}^\pm_{j+\frac{1}{8},m\pm\frac{9}{8}}.
\EN
This implies an isomorphism of the BRST cohomology classes
\EQ
H^n(\D^\mp_{jm})=H^n({\tilde\D}^\pm_{j+\frac{1}{8},
m\pm\frac{9}{8}}),
\EN
since $S^\pm$ commute with $Q_B$.

{}From the long exact sequence associated with the short exact
sequence (3.5) and the facts that $H^n({\cal F})=0$ for $n\ge 3$
and $n\le 0$ in the region $k,l\ge 0$ \cite{LIZ}, we find
\EQ
H^{n+1}({\tilde\D}^\pm_{jm})=H^n(\D^\mp_{jm}),
\EN
for $n \geq 3$ and $n \leq -1$. Using eq. (3.9), one finds
\bea
H^{n+1}({\tilde\D}^\pm_{jm})&=&
H^n({\tilde\D}^\pm_{j+\frac{1}{8},m\pm\frac{9}{8}}), \nonumber \\
H^{n+1}(\D^\mp_{j-\frac{1}{8},m\mp\frac{9}{8}})&=&
H^n(\D^\mp_{jm}),
\ena
for $n \geq 3$ and $n \leq -1$. Iterating (3.11) for $n \geq 3$, we get
\EQ
H^n({\tilde\D}^\pm_{jm})=
H^{n+q}({\tilde\D}^\pm_{j-\frac{q}{8},m\mp\frac{9}{8}q})
\EN
for any positive integer $q$. On the other hand, a state with $j,m$
and level $N(>0)$ can be nontrivial only if it satisfies the on-shell
condition. Due to eq. (3.2), any state representing
the RHS in (3.12) must be at the level
\bea
N= \left( 2j-\frac{q}{4}-1\right) ^2 - \frac{4}{9}\left( m \mp
 \frac{9}{8}q \right) ^2 \nonumber \\
\to -\frac{8}{16}q^2 < 0 \;\;\; {\rm for} \;\;\;  q \to \infty.
\ena
Obviously any state in the module cannot be on-shell
for sufficiently large $q$. We thus conclude the cohomology is trivial
for $n \geq 3$:
\EQ
H^n({\tilde\D}^\pm_{jm})=0 \;\;\; {\rm for} \;\; n \geq 3.
\EN
For $n \leq -1$, we may repeat the same argument into the large negative
$n$ and show
\EQ
H^n({\tilde\D}^\pm_{jm})=0 \;\;\; {\rm for} \;\; n \leq 0.
\EN
The same conclusion is valid for the modules $\D^\pm_{jm}$.
So the nontrivial states can exist only for $n=1$ and $2$ .

To determine $H^{1,2}$, we will start from points on the line $l=0$
where $S^\pm$ are well defined, then move up to larger $l$.
Consider states at $(k,l)=(2p-1, 0)$ for positive integer
$p$ and write the long exact sequence associated with (3.5)
\EQ
\setlength{\unitlength}{1mm}
\begin{picture}(40,37)
\put(30,35){0}
\put(35,30){$\nwarrow$}
\put(0,25){$H^2({\tilde\D}^+)  \to H^2({\cal F}) \to
H^2(\D^-)$}
\put(5,20){$\nwarrow$}
\put(9,19){\line(1,0){25}}
\put(38,15){\line(-1,1){4}}
\put(0,10){$H^1({\tilde\D}^+) \to H^1({\cal F})
 \to H^1(\D^-)$}
\put(10,5){$\nwarrow$}
\put(15,0){$0.$}
\end{picture}
\EN
We know $H^2({\cal F})=0$ and $H^1({\cal F})={\bf C}$
on the line $l=0$~\cite{LIZ}. Also from eq. (3.9), we have
$H^{1,2}(\D^-_{jm})
 = H^{1,2}({\tilde\D}^+_{j+\frac{1}{8},m+\frac{9}{8}})$.
Using this relation to the cohomology classes on the RHS of (3.16),
we find
\EQ
H^2(\D^-)= H^1(\D^-)= 0,
\EN
since the states in ${\tilde\D}^+_{j+\frac{1}{8},m+\frac{9}{8}}$
have $k>0$ but $l<0$ and thus cannot satisfy the on-shell condition
(3.3). Substituting these into (3.16), we find
\EQ
H^2({\tilde\D}^+) = 0, \;\;
H^1({\tilde\D}^+)={\bf C}.
\EN

Similarly at $(k,l) = (p-\shalf, 0)$, we have
\EQ
\setlength{\unitlength}{1mm}
\begin{picture}(40,36)
\put(30,35){0}
\put(35,30){$\nwarrow$}
\put(0,25){$H^2({\tilde\D}^-)  \to 0 \to H^2(\D^+)$}
\put(5,20){$\nwarrow$}
\put(9,19){\line(1,0){25}}
\put(38,15){\line(-1,1){4}}
\put(0,10){$H^1({\tilde\D}^-) \to {\bf C} \to H^1(\D^+)$}
\put(10,5){$\nwarrow$}
\put(15,0){$0.$}
\end{picture}
\EN
In this case, we can relate the cohomology on the left to that of
$\D^+$ with $l<0$ by using the action of $S^-$.
We find from (3.9)
\EQ
H^2({\tilde\D}^-) = H^1({\tilde\D}^-)=0,
\EN
which, when substituted into (3.19), tells us
\EQ
H^2(\D^+) = 0, \;\; H^1(\D^+)={\bf C}.
\EN

By repeating this procedure, moving up to larger $l$ and using the
fact that $H^2({\cal F})=H^1({\cal F})={\bf C}$ for $k,l>0$\cite{LIZ},
we can determine all the cohomology $H^{1,2}(\D^\pm)$ and $H^{1,2}
({\tilde\D}^\pm)$. On the points where $S^+$ ($S^-$) can act, the
cohomology $H^{1,2}(\D^\pm)$ ($H^{1,2}({\tilde\D}^\mp)$) on the right
(left) of the long exact sequence may be obtained from $H^{1,2}
({\tilde\D}^\pm)$ ($H^{1,2}(\D^\mp)$) with smaller $l$ by (3.9).
Then the cohomology on the left (right) $H^{1,2}({\tilde\D}^\mp)$
($H^{1,2}(\D^\pm)$) can be found from the long exact sequence.

For the points with both $k$ and $l$ positive odd integers, further
study is necessary since the modules $\D^\pm$ split into
${\tilde\D}^\pm$ and ${\cal U}$. This splitting yields the
short exact sequence
\EQ
0 \to {\tilde\D}^\pm \to \D^\pm \to {\cal U} \to 0.
\EN
The long exact sequences associated with
(3.22) tell us nontrivial cohomology classes $H^n({\cal U})$.

For continuous reps. ${\cal C}$ which appear on the points
$k,l\in 2{\bf Z}_+$, there is no difference between Fock and PF modules.
The cohomology coincides with that of the free boson modules \cite{LIZ}.

Proceeding this way, we may determine all the nontrivial cohomology
in the region $k,l>0$:
\bea
{\tilde\D}^+ : && H^2={\bf C} \;\; \qquad{\rm at} \;\;
 (k,l)=(2p,q-\shalf), \nonumber\\
 && H^1={\bf C} \;\; \qquad{\rm at} \;\; (k,l)=(2p-1,q),\nonumber\\
\D^- : && H^2={\bf C} \;\; \qquad{\rm at} \;\;
 (k,l)=(2p-1,2q), \nonumber \\
 && H^1={\bf C} \;\; \qquad{\rm at} \;\; (k,l)=(2p,q-\shalf),\nonumber\\
{\tilde\D}^- : && H^2={\bf C} \;\; \qquad{\rm at} \;\;
 (k,l)=(p-\shalf , 2q), \nonumber\\
 && H^1={\bf C} \;\; \qquad{\rm at} \;\; (k,l)=(p, 2q-1),\nonumber\\
\D^+ : && H^2={\bf C} \;\; \qquad{\rm at} \;\;
 (k,l)=(2p,2q-1), \nonumber \\
 && H^1={\bf C} \;\; \qquad{\rm at} \;\; (k,l)=(p-\shalf,2q),\nonumber\\
{\cal U} : && H^2 = H^0 ={\bf C} \;\; \qquad{\rm at} \;\;
 (k,l)=(2p-1,2q-1), \nonumber\\
{\cal C} : && H^2 = H^1 ={\bf C} \;\; \qquad{\rm at} \;\; (k,l)=(2p,2q),
\ena
for all positive integers $p$ and $q$. Others are trivial. The results
agree with those in ref.~\cite{DN} and are depicted in Fig. 2.

In the above analysis, we neglected the presence of null states in
the PF modules: they could be reducible.  We will see how the results
are modified when null states are taken into account.

\sect{Null states in PF modules}
\indent
We would like to identify null states in PF modules.
A PF module may be obtained from an $SL(2,R)$ current algebra module
by the condition $J^3 \sim 0$.  Thus null states in $SL(2,R)$
current algebra modules are always in PF modules; there is actually
a one-to-one correspondence between null states in two modules.
So it is sufficient to find null states in $SL(2,R)$ modules.

In this section, we first describe some general properties of
modules by using currents rather than the free field realization.
Next we construct null states with charge screening operators
by the procedure to be explained shortly. The arguments are generic and
the level of the algebra may take arbitrary value.
Some special features for $K=9/4$ will be used in the second part of
this section. These will be explained in some examples.

For concreteness let us take a LW rep. as a base rep., whose elements
are written as $|j,m\ra$ with LW state as $|j,j\ra$. The LW state is
characterized by the primary conditions: $J_0^-|j,j\ra=0$ and
$J_n^a|j,j\ra=0, (a=\pm, 3$ and $n>0)$. A rep. of the algebra is
generated by the action of $J_0^+$ and $J_{-n}^a, (n>0)$ on $|j,j\ra$.
Among the states in the generated module, we may have states which
satisfy the same condition as the LW state, $J_0^-|\chi\ra=0$ and
$J_n^a|\chi\ra=0, (a=\pm, 3$ and $n>0)$. These are called
null states and are to be removed to have an irrep.

It is well known that the algebra (2.9) is invariant under the
Weyl transformation
\EQ
J_n^{\pm} \to {\tilde J}_{n \mp 1}^{\mp}, \qquad
J_n^{3} \to -{\tilde J}_n^3 + {K \over 2}\delta_{n,0}.
\EN
Under this transformation, the rep. is rearranged into
another LW rep. From the following relations
\EQ
{\tilde J}_0^3 |j,m \ra = ({K \over 2}-m) |j,m\ra, \qquad
{\tilde {\bf J}}^2 |j,j \ra
= ({K \over 2}-j)(1-{K \over 2}+j) |j,j\ra,
\EN
we read changes in quantum numbers
\EQ
m \to {\tilde m}=K/2-m, \qquad
-j(j-1) \to ({K \over 2}-j)(1-{K \over 2}+j).
\EN
The original rep. may contain a null state
$|\chi\ra \equiv {\cal O} (J_0^+, J_{-n}^a) |j,j\ra$.
Since the algebra does not change under the Weyl transformation,
the state obtained from $|\chi\ra$ by the replacement (4.1) is a null
state in the transformed rep. In other words, the null states in two
reps. are related by the Weyl transformation.

Turning to the free field realization, a general formula for null states
is written in terms of charge screening operators. For the $SL(2,R)$
currents in eq.~(2.11), the screening operator is given by
\EQ
W(z) = \left(\sqrt{\frac{K-2}{2}}\pa\phi^L
 + i\sqrt{\frac{K}{2}}\pa\phi^M \right) \exp \left(\sqrt{\frac{2}{K-2}}
 \phi^L\right).
\EN
It is easy to see that $W$ obeys expected OPEs
\EQ
J^{\pm,3}(z)W(w) \sim \partial_w ({\rm something}).
\EN

We define singular vertex operators $\Phi_{jm}^s$ corresponding to
null fields by\footnote{The integrations are defined by the analytic
continuation in the level $K$.}
\bea
\Phi_{jm}^s (z)
&=& \oint \frac{dz_s}{2\pi i} \int_z^{z_s} \frac{dz_{s-1}}
 {2\pi i} \cdots \int_z^{z_2}\frac{dz_1}{2\pi i}
 W(z_s)W(z_{s-1}) \cdots W(z_1)U_{jm}(z) \nonumber\\
&\equiv& Q^s U_{jm}(z).
\ena
Since the screening charge (4.4) does not depend on $J^3$ or $\phi^3$,
these null states in current algebra modules are also
the null states in PF Verma modules, which are of our interest.

Let us find conditions that $\Phi_{jm}^s$ is well-defined and
non-vanishing. If we normal order the integrand in (4.6), we find that
the most singular part has the factor
\EQ
\prod_{i>j}^s (z_i - z_j)^{-2/(K-2)}
\prod_{i=1}^s (z_i - z)^{-2j/(K-2)-1}.
\EN
Thus the contour integral over $z_s$ is dragging the integrations over
$z_i, (i=1,2, \cdots, s-1)$. By changing the integration variables
as $z_i \to z_i z_s$, we see that the $z_s$ integration is
well-defined and non-vanishing when
\EQ
(s-1)-\frac{2}{K-2}\frac{s(s-1)}{2}-\frac{2js}{K-2}-s=-N-1,
\EN
where $N$ is a non-negative integer, which will be found to be the level
of the null state. Solving (4.8) for $j$ yields
\EQ
j=\frac{K-2}{2}\frac{N}{s}+\frac{1-s}{2}.
\EN
This is only a necessary condition for the presence of singular
vertex operators (4.6). As in the Virasoro null states~\cite{FEL,KMA},
the necessary and sufficient condition is actually that $\frac{N}{s}=r$
is a non-negative integer. Thus there exists a null state for
$U_{jm}$ with $j=j_{(r,-s)}$, where
\EQ
j_{(r,s)} \equiv \frac{K-2}{2}r+\frac{1+s}{2}.
\EN
Since $W$ carries $j$ quantum number, the Casimirs for $\Phi_{jm}^s$ and
$U_{jm}$ are given by ${\bf J}^2=-j(j-1)$ with $j=j_{(r,s)}$ and
$j=j_{(r,-s)}$, respectively; the map does not change $m$.
Therefore the level of a null state (cf. (3.2)) is
\EQ
\frac{1}{K-2}\left[ j_{(r,s)}(j_{(r,s)}-1)
 -j_{(r,-s)}(j_{(r,-s)}-1) \right] = rs=N.
\EN

The primary conditions on singular vertex operators are reduced to
the same conditions on $U_{jm}$ due to (4.5).
Thus $U_{jm}$ itself must correspond to a LW (HW) state for
$\Phi_{jm}^s$ to be null vectors. This gives us a condition on $m$
as $m \mp j=0$.

Proceeding as in the Virasoro case~\cite{KMA}, we are led to the
Kac-Kazhdan formula~\cite{KK,DLP}
\EQ
D_N = (-1)^{r_3(N)}C_N(K-2)^{r_3(N)}
 \prod_{\stackrel{r,s=1}{rs\leq N}}^N \left[ -j(j-1)
 +j_{(r,s)}(j_{(r,s)}-1) \right]^{p_3(N-rs)},
\EN
for the determinant of inner products of all states with fixed $m$
at level $N$. Here $C_N$ is a numerical constant.
The exponent $r_3(n)$ is the number of factors of $J^3_{-p}$
in all states at level $n$ in a system with three generators ;
$p_3(n)$ is the number of states at level $n$
in a theory of three free bosons.

The complete degenerate reps. of $SL(2,R)$ current algebra
are clarified in refs.~\cite{BF,KY}. The case of our interest
corresponds to subcase F in the classifications~\cite{KY} and
primary states are on the boundary of a conformal grid.
Indeed in our model of $SL(2,R)_{9/4}$, the conformal grid has
the size $1\times 4$ and all the states are on its boundary.

A characteristic feature of these boundary modules is that
the embedding structure of null states is quite simple;
null states at the higher levels belong to submodules generated
from null states at lower levels. Therefore in order to get
an irrep., it is enough to subtract
the submodule generated from the lowest level null state.

The embedding structure of null states is a little complicated
when there is a ${\cal U}$ rep. since we have two
null states at the same level in this case. Let us explain this
for $K=\frac{9}{4}$,
in a simple example, by using the above consideration on
the Weyl transformation and the construction of null states by free
fields. Consider a current algebra module on the base with $m=
17/8, 17/8+1, \cdots$ in Fig. 3(a). There are three null states in
this module at levels 3, 9 and 10 as dictated by the determinant formula
(4.12). This may be understood in the free field realization as well:
since the Casimir corresponds to $j=17/8$, we may choose $(r,s)=(1,3),
(5,2)$ and $(9,1)$. Therefore we may construct singular vertex
operators associated with $(r,-s)=(1,-3), (5,-2)$ and $(9,-1)$ at
levels 3, 10 and 9. Null states at higher levels are contained in
modules generated from those at lower levels~\cite{KY}.

After Weyl transformation, this module becomes that
over the state indicated as $\otimes$ in Fig. 3(b).
We have one null state at level 0 and other two null
states at the same level 8. Each pair of null states with the same
symbols in the figures are related by the transformation.

This new module is $\D^+$ with $j=2$ or $-j(j-1)=-2$, for which we may
construct three null states associated with $(r,-s)=(0,-3), (4,-2)$
and $(8,-1)$. The null state mapped from $(0,-3)$ has level $N=0$
and it decomposes this LW module to ${\cal U}$ and $\tilde{\D}^+$
modules. The null states with ${\tilde m}=0$ and 1 is mapped from
$(4,-2)$ and $(8,-1)$, respectively.

Since the embedding structure does not change under the Weyl
transformation, we can conclude that ${\cal U}$ module has no null state.
In addition, the null state with ${\tilde m}=1$ is obtained from that
with ${\tilde m}=0$ by the action of ${\tilde J}_0^+$.
We only need to remove the module generated
from the latter in order to obtain a $\tilde{\D}^+$ irrep.

Some explanations on $N=0$ null state are in order. The parent
$j_{(0,-3)}$ module related to this null state has a LW state with
${\tilde m}=-1$ rather than ${\tilde m}=2$. It thus appears that it
gives a null at ${\tilde m}=-1$. In this $N=0$ case, however, the map
of screening charge (4.6) is special because
only the most singular part contributes to the integral and
its coefficient is proportional to ${\tilde m}({\tilde m}
^2-1)$. Thus the state ${\tilde m}=2$ in $j_{(0,-3)}$ module
consistently creates the null state with ${\tilde m}=2$ in $j=2$ module.

These embedding structures are common in any $\D^+ (= {\cal U}
\oplus {\tilde {\D}^+})$ with $m\pm j\in {\bf Z}$.

In $\D^+$ there is a null state at level zero which separates the module
into ${\cal U}$ and ${\tilde {\D}^+}$ modules.
All the higher level null states are in ${\tilde {\D}^+}$ module.
Thus there is no null state in ${\cal U}$ module.
There are two degenerate lowest null states in ${\tilde{\D}^+}$. The
module $\D^+$ over the null with smaller ${\tilde m}$ is separated into
${\cal U}$ and ${\tilde \D}^+$ due to another null with larger
${\tilde m}$. All the higher null states generate submodules in this
$\D^+$. Hence one can obtain the irrep. by subtracting this module
$\D^+={\cal U}\oplus{\tilde \D}^+$.

One can easily find a similar structure of null states in HW modules.

Finally we consider the embedding structure of null states
in ${\cal C}$ module.\footnote{
A null state in ${\cal C}$ is defined by the condition
$J^{\pm,3}_n\Phi^s_{jm}=0$ for $n>0$.}
Using the map of the screening charge (4.6),
we can formally construct two null states at the same level.
However, one of these maps
cannot be defined by simple analytic continuation.
It is not clear for us when this map is well-defined.
This leaves us with two possibilities:\footnote{
Similar indeterminancy of the number of null states
appears whenever $K-2=\frac{1}{n}$ with positive integer $n$.
We have explicitly checked that there is only one null state
at the level $N=2$ in the case $n=1$ ($K=3$). We thus expect the second
structure is the correct one.}

(1) There are two lowest null states in a module.

The embedding structure of this case is obtained by counting
the number of null states in all the submodules.
This is illustrated by the following diagrams
\bea
&&\begin{array}{cccccccccccccc}
 & & \nearrow & \bullet & \to   & \bullet & \to &
 \bullet & \cdots & \bullet & \to & \bullet & \searrow & \\
j={\rm half \ integer:\ }& \bullet &    &       & \uda  & & \uda & &
 \cdots & & \uda & & & \bullet \\
 & & \searrow & \bullet & \to   & \bullet & \to & \bullet &
 \cdots & \bullet & \to & \bullet & \nearrow & ,
\end{array} \nonumber\\
\vs{5}
&&\begin{array}{cccccccccccc}
 & & \nearrow &\bullet & \to & \bullet & \to & \bullet & \cdots &
 \bullet & \to & \bullet \\
j={\rm integer:\ }& \bullet & & & \uda & & \uda & &\cdots & &\uda & \\
 & & \searrow & \bullet & \to & \bullet & \to & \bullet & \cdots
 & \bullet & \to & \bullet ,
\end{array}
\ena
where $\bullet$ represent null states. The null state at the end of an
arrow
is in the module over that at the origin of the arrow.

(2)  There is only one lowest null state in a module.

The embedding structure of this case is the same as the LW or HW
modules. Null states at higher levels are in the module over the lower
null states.

\sect{Physical states for $SL(2,R)/U(1)$}
\indent
As explained in the previous section, once we identify the lowest null
state, we may subtract the submodule generated over it to get an irrep.

We have shown that there is a null state for
\EQ
j=\frac{1}{8}(r+4s+4); \;\; r,s \in {\bf Z}_+,
\EN
at the level $N=rs$ for $K=\frac{9}{4}$.
Since $j$ is invariant under the following change of $r$ and $s$:
\EQ
r \to r'\equiv r+4p, \;\; s \to s'\equiv s-p, \;\; p \in {\bf Z},
\EN
the module has several null states at levels
\EQ
N=r's'=(r+4p)(s-p).
\EN
Each null state is generated from the module at $(r',-s')$ with
screening operators $Q^{s'}$. It is easy to count the number of null
states in a module. Since the level (5.3) must be positive, we find
\EQ
s>p>-\frac{r}{4}.
\EN
Using the freedom (5.2), we may always choose $r$ in the range
$4 \geq r \geq 1$. It then follows from (5.4) that $p$ can take values
$p = 0, 1, \cdots, s-1$, which implies that there are $s$ null states.
{}From (5.1) and (3.1), $(k,l)$ and $(r,s)$ are related as $k+l=r/2+2s$.
This defines lines on the $(k,l)$ plane and multiplicities of null
states on them are $s$. On the $(k,l)$ plane in Fig. 4, we show the
number of null states at points studied in sect. 3.

Now let us compute the cohomology of irreps.
Consider the short exact sequence
\EQ
0 \to \D^{(n-1)}_{j(r,-s),m} \to \D^{(n)}_{j(r,s),m}
\to {\hat\D}_{j(r,s),m} \to 0,
\EN
where $\D^{(n)}_{jm}$ is the PF Verma module with $n$ null
states and ${\hat\D}_{jm}$ is an irrep., whose cohomology is of our
interest. The first map with $Q^s$ is the embedding of $\D^{(n)}_{jm}$
onto the submodule over {\em the lowest level} null state in
$\D^{(n)}_{jm}$. The cohomology for the modules $\D^{(n-1)}$ and
$\D^{(n)}$ will be obtained from our earlier results in sect. 3. The
cohomology for the irrep. ${\hat\D}$ is then
calculated from the long exact sequences associated with (5.5).

Let us explain our procedure in more detail. Take a point in the first
quadrant on the $(k,l)$ plane studied in sect. 3. From (3.1) and (5.1),
we find $2j_{(r,s)}-1=r/4+s=(k+l)/2$ and $2m/3=(k-l)/2$. We know
the cohomology for modules with $j_{(r,s)}$ and $m$.

In order to construct an irrep. ${\hat\D}_{j(r,s),m}$, we would like
to identify $\D^{(n-1)}_{j(r,-s),m}$ with less null states by one
for a given $\D^{(n)}_{j(r,s),m}$. Note that the two modules have
different $j$ but the same $m$. From this observation, we may find
$(k,l)$ coordinates for $\D^{(n-1)}_{j(r,-s),m}$, $(k',l')$, as follows:
$k'=k-2s$ and $l'=l-2s$. When $4\ge r \ge 1$, $Q^s$ realizes the
desired map from $\D^{(n-1)}_{j(r,-s),m}$ to $\D^{(n)}_{j(r,s),m}$
so that we have an irrep. ${\hat \D}$.
The changes in $(k,l)$ coordinates by the map are
\bea
(k',l')=(-2q,-2p + i/2) &\to& (2p, 2q + i/2 )=(k,l), \nonumber \\
(-2q + i/2,-2p) &\to& (2p + i/2, 2q ), \nonumber\\
(-2q,-2p) &\to& (2p+2, 2q+2), \nonumber\\
(1-2q,1-2p) &\to& (2p+1, 2q+1),
\ena
where $p, q$ are positive integers and $i = 1, 2, 3$. Since both
$k'$ and $l'$ are negative for $\D^{(n-1)}_{j(r,-s),m}$, we realize
that it is dual to some module in the first quadrant listed in Fig. 4.
When a module has quantum numbers $(j,m)$, the dual module carries
$(1-j,m)$.  This corresponds to the change $(k,l) \leftrightarrow
(-l,-k)$ on the $(k,l)$ plane. With the isomorphism in (2.26), the
cohomology at $(k',l')$ is obtained by the replacement
\EQ
{\tilde\D}^- \leftrightarrow \D^- ,
{\tilde\D}^+ \leftrightarrow \D^+,
H^2 \leftrightarrow H^0.
\EN
from the results for $(-l',-k')$ in sect.~3.
Here the third relation is due to the dual structure of ghost
Fock space.  It is useful to see relations between $(-l',-k')$ and
$(k,l)$ on Fig. 4.
Note that the points $(-l',-k')$ and $(k,l)$ are on the same line
$k-l=\frac{4}{3} m$ and the change of $k$ coordinate is $k \to
-l'=2s-l$. We see that
(1) when either $k$ or $l$ is a half-integer,
$(-l',-k')$ is the closest point to the left on the line;
(2) when both $k$ and $l$ are integers, $(-l',-k')$ is the
next closest point to the left on the line.
The change of $s$, the number of null states, is one for (1) and
two for (2).

As $\D^{(n)}_{j(r,-s),m}$, we study four different cases with $(k,l)=$
(i) $(2p, q-1/2)$, (ii)$(2p-1,2q)$, (iii)$(2p-1,2q-1)$ and (iv)$(2p,2q)$.
Other possibilities are realized by exchanging $k$ and $l$. The possible
patterns for long exact sequences to be studied later are given below:
\EQ
\setlength{\unitlength}{1mm}
\begin{picture}(90,45)(-20,0)
\put(-7,40){0}
\put(-6,35){$\nwarrow$}
\put(-18,30){$a \to {\bf C} \to H^2$}
\put(-17,25){$\nwarrow$}
\put(-13,24){\line(1,0){12}}
\put(3,20){\line(-1,1){4}}
\put(-18,15){${\bf C} \to b \to H^1$}
\put(-17,10){$\nwarrow$}
\put(-13,9){\line(1,0){12}}
\put(3,5){\line(-1,1){4}}
\put(-18,0){$0 \to 0 \to H^0$}
\put(-17,-5){$\nwarrow$}
\put(-13,-6){\line(1,0){12}}
\put(3,-10){\line(-1,1){4}}
\put(-8,-15){$0 \to H^{-1}$}
\put(-8,-25){(A)}
\put(38,40){0}
\put(40,35){$\nwarrow$}
\put(25,30){$0 \to 0 \to H^2$}
\put(26,25){$\nwarrow$}
\put(30,24){\line(1,0){12}}
\put(46,20){\line(-1,1){4}}
\put(25,15){$a \to {\bf C} \to H^1$}
\put(26,10){$\nwarrow$}
\put(30,9){\line(1,0){12}}
\put(46,5){\line(-1,1){4}}
\put(25,0){$b \to 0 \to H^0$}
\put(26,-5){$\nwarrow$}
\put(30,-6){\line(1,0){12}}
\put(46,-10){\line(-1,1){4}}
\put(35,-15){$0 \to H^{-1}$}
\put(37,-25){(B)}
\end{picture}
\EN
\vs{15}

We will indicate below the long exact sequence for each case,
from which the desired cohomology can be obtained.
The cohomology is summarized in eq. (5.10).

\noindent\underline{Case (i) $(k,l)=(2p, q-1/2)$}

{}From Fig. 2, we find that $H^2({\tilde \D}^+)= H^1(\D^-)={\bf C}$
for $\D^{(n)}$. When $q=1,2$, $l'=0$ and the only nontrivial cohomology
is $H^1(\D^+)={\bf C}$, corresponding to the tachyon.
Applying the rule (5.7), we find $H^1({\tilde \D}^+)={\bf C}$ for
$\D^{(n-1)}$. The long exact sequence for
${\hat {\tilde \D}}^+$ is in the pattern (A) with $a=b=0$.
It is easy to find that $H^{0,2}({\hat {\tilde \D}}^+)={\bf C}$.
For ${\hat \D}^-$, the sequence is in the pattern (B) with $a=b=0$ and
the result is  $H^{1}({\hat {\D}}^-)={\bf C}$.
When $q>2$, $H^2({\tilde \D}^-)=H^1(\D^+)={\bf C}$ for $(-l',-k')$, so
$H^0(\D^-)=H^1({\tilde \D^+})={\bf C}$ for $\D^{(n-1)}$ at $(k',l')$.
The sequences are (A) with $a=b=0$ for ${\hat {\tilde \D}}^+$ and
(B) with $a=0$ and $b={\bf C}$ for ${\hat \D}^-$.

\noindent\underline{Case (ii) $(k,l)=(2p-1, 2q)$}

We have $H^1({\tilde \D}^+)= H^2(\D^-)={\bf C}$ for $\D^{(n)}$.
We find $H^2(\D^+)=H^1({\tilde \D}^-)={\bf C}$ at $(-l',-k')$, so
$H^0({\tilde \D}^+)= H^1(\D^-)={\bf C}$ for $\D^{(n-1)}$. The sequences
are (A) with $a=b=0$ for ${\hat \D}^-$ and (B) with $a=0$ and $b={\bf C}$
for ${\hat {\tilde \D}}^+$. When $p=1$, we have to modify the above
argument slightly:  $H^0({\tilde \D}^+)=0$ for $\D^{(n-1)}$
and the pattern is (B) with $a=b=0$ for ${\hat {\tilde \D}}^+$.

For cases (iii) and (iv), there are two lowest null states associated
with two possible values for $(r, -s)$ (with $r=4$ or $s=1$). So other
than the mapping given in (5.6), there could be another map, for which
$(k,l)$ coordinates change as
\EQ
(k',l')=(p-2,q-2)\to(p,q)=(k,l)
\EN
Hence we need further investigation to obtain the cohomology of irreps.

\noindent\underline{Case (iii) $(k,l)=(2p-1, 2q-1)$}

For this case, we have already clarified the embedding structure of
the null states in the previous section. We should use (5.6)
to obtain the irreps.

Nontrivial cohomology classes for $\D^{(n)}$ are
$H^{2,0}({\cal U})=H^1({\tilde \D}^{\pm})={\bf C}$.
As we have shown in sect. 4, there is no null state in ${\cal U}$ reps.
so that the results in sect. 3 are valid for them. We only
have to study ${\tilde \D}^{\pm}$ reps. When $p=q=1$,
there is no null state and the module is irreducible.  When either
$p$ or $q$ is 1, say $(k,l)=(2p-1,1)$, $(k',l')=(1,3-2p)$ and the
cohomology for $\D^{(n-1)}$ is trivial. The long exact sequence is
(B) with $a=b=0$ for ${\hat {\tilde \D}}^{\pm}$.
For a generic $(p,q)$, we find $(k',l')=(3-2q, 3-2p)$.
The sequence is (B) with $a={\bf C}$ and $b=0$ for
${\hat{\tilde\D}^{\pm}}$.
Due to the pattern, $0 \to H^0 \to {\bf C} \to {\bf C}
\to H^{1} \to 0$, there are two possible solutions ;
$H^{1,0}(\hat{\tilde \D}^{\pm})=0$ or ${\bf C}$.
Accordingly we find two possibilities
for the cohomology for ${\hat {\tilde \D}^{\pm}}$.
The first possibility is shown in Fig. 5.

\noindent\underline{Case (iv) $(k,l)=(2p, 2q)$}

Here we have only ${\cal C}$ modules. It is crucial to know the
embedding structure of null states since two possibilities
of $\D^{(n-1)}$ in (5.6) and (5.9) give different cohomologies.
We have the following two possibilities for the structure.

(1) There are two null states in a module: the above two maps
give different null states.

(2) There is only one null state in a module: the map
$(2p-2,2q-2)\to(2p,2q)$ give the null state and the map
$(2-2q,2-2p)\to(2p,2q)$ is ill-defined.\footnote{
The map $(2p-2,2q-2)\to(2p,2q)$ is always well-defined
since it has only one integration with a closed contour.}

Let us compute the cohomology of irreps. for each case.

For the first possibility,
the embedding structure of null states is not so simple,
as explained in the previous section. It is not enough to
subtract modules over the lowest level null states;
we have to subtract irreps. over all the null states in order to
get the right cohomology of an irrep. Note, however, that if we know the
cohomology of irreps. with less null states, we can find
the cohomology of a module by subtracting the irreps. Since submodules
on the points with less $k+l$ (on the $(k,l)$ plane) have less null
states, we can compute the cohomology of irrep. inductively, starting
from modules at smaller $k+l$ to those at larger $k+l$.
In this way, we can compute all the cohomology of irreps. in principle.
However, the long exact sequence does not give
a unique cohomology as in the case (iii). Since the cohomology of
an irrep. is obtained by using that with less null states, we have
more possibilities for the former cohomology than the latter.

For the second possibility,
we find $H^{2,1}({\cal C})={\bf C}$ for $\D^{(n)}$. When $p=1$ or $q=1$,
we have $H^1={\bf C}$ for $\D^{(n-1)}$. The pattern is (A) with $a=0$
and $b={\bf C}$ and $H^2({\hat{\cal C}})={\bf C}$.
As for $H^{1,0}({\hat {\cal C}})$, we encounter the same indeterminancy
mentioned for Case (iii), obtaining $H^{1,0}(\hat{\cal C})=0$ or {\bf C}.
Fig. 5 shows the latter possibility.
When $p,q > 1$, $H^1({\cal C})=H^2({\cal C})={\bf C}$
at $(k',l')=(2p-2,2q-2)$ for $\D^{(n-1)}$.
The long exact sequence is (A) with $a=b={\bf C}$.
We have four possibilities for the cohomology:
$(H^0,H^1,H^2)=(0,0,0),\ (0,{\bf C},{\bf C}),\ ({\bf C},{\bf C},0)$ or
$({\bf C},{\bf C}\oplus{\bf C},{\bf C})$.
The second possibility is shown in Fig. 5.

Let us summarize the nontrivial cohomology obtained in our study:
\bea
{\hat{\tilde\D}}^+ : &&
H^2=H^0={\bf C} \qquad{\rm at}\qquad (2p, q-\shalf), \nonumber \\
&& H^1={\bf C} \qquad{\rm at} \qquad(2p-1,2q),\; (2p-1,1),\;(1,2q-1),
\nonumber\\
&& H^1=H^0=0\ {\rm or}\ {\bf C} \qquad{\rm at}\qquad(2p+1,2q+1)
\nonumber\\
&& H^{-1}={\bf C} \qquad{\rm at} \qquad(2p+1, 2q),\nonumber\\
{\hat\D}^- : && H^2=H^0={\bf C} \qquad{\rm at}\qquad
 (2p-1, 2q), \nonumber \\
&& H^1={\bf C} \qquad{\rm at}\qquad (2p, q-\shalf), \nonumber\\
&& H^{-1}={\bf C} \qquad{\rm at}\qquad (2p, q+\frac{3}{2}),\nonumber\\
{\hat{\tilde\D}}^- : &&
H^2=H^0={\bf C} \qquad{\rm at}\qquad (p-\shalf, 2q), \nonumber \\
&& H^1={\bf C} \qquad{\rm at}\qquad (2p, 2q-1),\; (2p-1,1),\; (1,2q-1),
 \nonumber\\
&& H^1=H^0=0\ {\rm or}\ {\bf C} \qquad{\rm at}\qquad(2p+1,2q+1)
\nonumber\\
&& H^{-1}={\bf C} \qquad{\rm at}\qquad (2p, 2q+1),\nonumber\\
{\hat\D}^+ : &&
H^2=H^0={\bf C} \qquad{\rm at}\qquad (2p, 2q-1), \nonumber \\
&& H^1={\bf C} \qquad{\rm at}\qquad (p-\shalf, 2q), \nonumber\\
&& H^{-1}={\bf C} \qquad{\rm at}\qquad (p+\frac{3}{2}, 2q),\nonumber\\
{\hat{\cal U}} : &&
H^2 = H^0 ={\bf C} \;\; \qquad{\rm at}\qquad
(2p-1,2q-1),\nonumber\\
{\hat{\cal C}} : &&
(H^2,H^1,H^0)=(0,0,0) \quad{\rm or}\quad (0,{\bf C},{\bf C})
\quad{\rm or} \quad ({\bf C},{\bf C},0)
\quad{\rm or}\quad ({\bf C},{\bf C}\oplus{\bf C},{\bf C})
\nonumber\\ && \qquad\qquad\qquad\qquad{\rm at}\qquad (2p+2, 2q+2),
\nonumber\\
&&H^2={\bf C}, H^1=H^0=0 \quad{\rm or}\quad {\bf C}\qquad{\rm at}
\qquad (2, 2q),\ (2p,2),
\ena
for all positive integers $p$ and $q$. Here we only list the results
of case (2) for ${\cal C}$ rep. There is some indeterminancy in
our results at the points $(2p,2q)$ and $(2p+1, 2q+1)$.
We have shown in Fig. 5 one possibility of these results.

Now let us compare this spectrum with that of $c=1$ gravity. The
discrete states of $c=1$ gravity correspond to states with $k,l
\in {\bf Z}_+$ and ghost numbers $N_{FP}=1, 2$. Our
maximum spectrum includes these states. In addition to these states,
there are an infinite number of discrete states in the spectrum. We give
a simple example to show how these physical states appear in the next
section.

\sect{Discussions}
\indent
Using the cohomological terms, we have identified the nontrivial
cohomology in the $SL(2,R)/U(1)$ coset model. The physical states in
the theory exist for the ghost number $N_{FP}=-1\sim 2$.

Let us now discuss the character formulae for reps. of the
coset model and their relation to our results. As discussed by Distler
and Nelson~\cite{DN}, if we define the character-valued index by
\EQ
{\rm Ind}(q) =  q^{-1}\prod_{n=1}^{\infty} (1-q^n)^2 \chi_{jm}(q),
\EN
where $\chi_{jm}(q)$ is the character Tr$(q^{L_0})$ of the
$SL(2,R)/U(1)$ PF module, the $q^0$ term in Ind$(q)$ gives
the index of the BRST operator
\EQ
{\rm Index}(Q_B)= \sum_n (-1)^{n+1} {\rm dim}(H^n).
\EN
Using the character formulae, Distler and Nelson computed the index,
the alternating sum of the dimensions of the cohomology. Let us see
how this works.

The characters for hermitian reps. are known as~\cite{CHA,DN}
\bea
\chi_{jm}({\cal C}) &=&  q^{h_{jm}}\prod_{n=1}^\infty (1-q^n)^{-2},
\nonumber \\
\chi_{jm}({\tilde\D}^-) &=& q^{h_{jm}}\prod_{n=1}^\infty
 (1-q^n)^{-2} \sum_{s=0}^\infty (-1)^s q^{s(s-2m-2j+1)/2}, \nonumber \\
\chi_{jm}({\tilde\D}^+) &=& \chi_{j,-m}({\tilde\D}^-),
 \nonumber \\
\chi_{jm}({\cal D^-}) &=& \chi_{1-j,m}({\tilde\D}^-),
 \nonumber \\
\chi_{jm}({\cal D^+}) &=& \chi_{1-j,-m}({\tilde\D}^-), \nonumber \\
\chi_{jm}({\cal U}) &=& \prod_{n=1}^\infty (1-q^n)^{-2}
 \left[ 1- \sum_{s=0}^\infty (-1)^s q^{s(s-2j-1)/2}(q^{ms}+q^{-ms})
 \right].
\ena
Actually these characters must be modified since there exist null
states in general. If we use them without modifications, they give
the indices corresponding to the cohomology over the PF Verma modules.
Let us show that our results obtained in sect. 3 are consistent with
these indices.

For ${\tilde\D}^-$ reps., using (3.2) and (3.3),
we see that the condition that the index of $Q_B$ is not zero gives
\EQ
-kl + \shalf s(s-2k+l)=0,
\EN
for non-negative integer $s$. This has the solution
\EQ
k=\frac{s}{2}.
\EN
Combined with the condition $m-j=\shalf (k-2l-1) =$ integer, this
gives the following solutions. If $s$ is even $s=2{\tilde s}$,
the index is $+1$ for
\EQ
k={\tilde s}, \;\; l=2r+1; \;\; r, {\tilde s} \geq 0.
\EN
If $s$ is odd $s=2{\tilde s}+1$, the index is $-1$ for
\EQ
k={\tilde s}+\shalf , \;\; l=2r; \;\; r, {\tilde s} \geq 0.
\EN
This result is consistent with that in sect. 3.

It is easy to check that for all other reps. this analysis
gives results consistent with those in sect. 3, although it does not
determine precisely where the physical states are. Moreover, it cannot
determine whether there are physical states when the index is zero.

The characters of the irreps. are obtained by
subtracting the contributions of null states.
Let us discuss how the above analysis is modified when this point
is taken into account. In order to derive the correct characters,
note that it is enough to subtract submodule generated on the null
states which are indicated in Fig. 4.
For example, consider the $\tilde \D^+$ rep. with
$j=\frac{1}{4}(2p+2q-\frac{1}{2}+2)$ for positive integers $p,q$.
The contributions of the null states are subtracted if we subtract
the character at $j'=-j+\frac{7}{4}$. We have the character
\EQ
{\hat\chi}_{jm}(\hat{\tilde{\cal D}}^+) =
\chi_{jm}(\tilde{\cal D}^+)-\chi_{-j+7/4,m}(\tilde{\cal D}^+)
\EN
The same analysis as above then tells us that the nonvanishing index is
\EQ
{\rm Index}(Q_B)=-2,
\EN
in agreement with the results in sect. 5.
At more general points where the cohomology is nontrivial,
we get the characters of irreps.
\EQ
{\hat \chi}_{jm} = \chi_{jm}-\chi_{-j+(8-i)/4,m},
\EN
for $(k,l)=(2p,2q-\frac{i}{2})$ or $(2p-\frac{i}{2},2q)$ with $i=1,2,3$.
For the points $(k,l)=(2p+1,2q+1)$, the null state structure is a bit
complicated as shown in sect. 4. There is no null state in ${\cal U}$
reps. and the complementary $\tilde{\cal D}^\pm$ reps. have two null
states at the lowest level. Thus we get for these reps.
\bea
{\hat \chi}_{jm}(\hat{{\cal U}}) &=& \chi_{jm}({\cal U}),\\
{\hat \chi}_{jm}(\hat{\tilde{\cal D}}^\pm) &=&
\chi_{jm}(\tilde{\cal D}^\pm)-\chi_{2-j,m}({\cal U})
-\chi_{j-1,m}(\tilde{\cal D}^\pm).
\ena
For ${\cal C}$ reps., we obtain two possible characters corresponding
to the cases given in the previous section. For case (1), the character
is given by an alternative sum as usual complete degenerate reps.
Here we give it as a recursion relation
\EQ
\hat{\chi}_{jm}(\hat{{\cal C}})
=\chi_{jm}({\cal C})-\chi_{2-j,m}({\cal C})
-\hat{\chi}_{j-1,m}({\hat{\cal C}}),
\EN
where the initial characters are
\bea
\hat{\chi}_{\frac{3}{2},m}
&=&\chi_{\frac{3}{2},m}-\chi_{\frac{1}{2},m},\nonumber\\
\hat{\chi}_{2,m}
&=&\chi_{2,m}-\chi_{0,m}-\chi_{1,m},
\ena
for $j=$ half-integer and integer, respectively.
For case (2), the character has the similar form to other reps.
It is simply obtained by single subtraction:
\EQ
\hat{\chi}_{jm}(\hat{{\cal C}})
=\chi_{jm}({\cal C})-\chi_{j-1,m}({\cal C}).
\EN
One can see that these characters give indices
consistent with our results in sect. 5.

In this paper, we have considered only holomorphic sector of the theory.
For constructing a complete theory of the string,
we must take tensor products of holomorphic and antiholomorphic sectors
so as to give modular invariant partition function.
The complete classification of the modular invariant combinations
of the characters is left to future investigation.

We have proved that there are an infinite number of
discrete states in the string theory obtained from the $SL(2,R)/U(1)$
coset model. These discrete states include additional states
which are not present in the $c=1$ gravity. In ref.~\cite{EY},
however, it has been claimed that the spectrum is isomorphic
to that of $c=1$ gravity and there is no additional physical state
in contrast to our results.
Let us discuss why there is such a discrepancy
and how our new physical states appear by a simple example.

Our example is a set of the states with $j=m=9/8$ at level 1,
which correspond to the additional discrete states absent from the
spectrum of the $c=1$ gravity (Fig. 6). We first examine the BRST
quartet structure in
the vector space ${\cal F}^+_j\otimes{\cal F}_{gh}$,
where ${\cal F}^+_j$ is defined in sect. 2 and ${\cal F}_{gh}$
is the ghost Fock space. The vector space ${\cal F}^+_{9/8}$ has
two on-shell states at level $N=1$:
\bea
|\psi_1\ra&=&J^+_{-1}|\frac{9}{8},\frac{1}{8}\ra
+\frac{5}{4}J^-_{-1}|\frac{9}{8},\frac{17}{8}\ra, \nonumber\\
|\psi_2\ra&=&J^-_{-1}|\frac{9}{8},\frac{17}{8}\ra
-\frac{8}{9}J^3_{-1}|\frac{9}{8},\frac{9}{8}\ra,
\ena
which satisfy the condition
\EQ
J^3_n|\psi_{1,2}\ra=0,\quad n\ge 1.
\EN
We find the following structure under the action of the BRST charge on
the states:
\bea
Q_Bb_{-1}||\frac{9}{8},\frac{9}{8}\ra\ra
&=&9||\psi_2\ra\ra,\\
Q_B||\psi_1\ra\ra
&=&\frac{5}{2}c_{-1}||\frac{9}{8},\frac{9}{8}\ra\ra,
\ena
where we denote the tensor product state with ghost physical vacuum as
$||*\ra\ra=|*\ra\otimes c_1|0\ra$. Thus all the states in (6.18) and
(6.19) are in the BRST quartet and decouple from the physical subspace
in the vector space ${\cal F}^+_j\otimes{\cal F}_{gh}$.

In the PF irrep. (with the ghost Fock space), however, the BRST quartet
structure is different. First of all, it is true that there is a state
which has nonzero inner product with $|\psi_2\ra$ {\em in the vector
space} ${\cal F}^+_j$, but $|\psi_2\ra$ is a null state {\em in
the PF module} and must be set equal to zero to obtain the PF irrep.;
$b_{-1}||\frac{9}{8},\frac{9}{8}\ra\ra$ is a singlet in the PF rep.

To discuss the second doublet relation (6.19), we must first explain
how the space of base states $F^{\pm}_j=\oplus_{m=\pm j+{\bf Z}}F_{j,m}$
(see (2.22)) splits into the base reps. As explained in sect. 2,
the space $F^{\pm}_j$ splits into two irreps. as $F^{\pm}_j
={\tilde\Delta}^{\pm} \oplus\Delta^{\mp}$ when $j\mp m\in {\bf Z}$.
However, the transformations by currents do not respect this property
in the whole space $F^{\pm}_j$. From the OPE (2.14), we obtain
\bea
J^{\pm}_0|j,m\ra&=&(m\pm j)|j,m\pm 1\ra, \nonumber\\
J^3_0|j,m\ra&=&m|j,m\ra,
\ena
which shows that $J^-_0$ correctly annihilates the lowest state
$|j,j\ra$ in ${\tilde \Delta}^+$:
\EQ
J^-_0|j,j\ra=0,
\EN
but $J^+_0$ does not the highest $|j,j-1\ra$ in $\Delta^-$:
\EQ
J^+_0|j,j-1\ra=(2j-1)|j,j\ra\ne 0,
\EN
which must be put to zero.
Thus, for the whole PF modules, the states in
${\tilde\D}^{\mp}$ are null if they appear upon applying the PF
currents on $\D^{\pm}$, as defined in eq. (2.23).

In the above example, we immediately notice that the state $|\psi_1\ra$
does not make sense as a PF rep. since it consists of two
states $J^+_{-1}|\frac{9}{8},\frac{1}{8}\ra$ and $J^-_{-1}|\frac{9}{8},
\frac{17}{8}\ra$ in different reps. $\D^-$ and
${\tilde\D}^+$, respectively (see Fig. 6). We must
reinterpret these states as the PF reps.

For the state in $\D^-$, we take
$|\psi'_1\ra=J^+_{-1}|\frac{9}{8},\frac{1}{8}\ra$.
At first sight, this state does not appear to be in the PF module since
\EQ
J^3_1|\psi'_1\ra
=J^+_0|\frac{9}{8},\frac{1}{8}\ra
=\frac{5}{4}|\frac{9}{8},\frac{9}{8}\ra,
\EN
which is responsible for the doublet structure in (6.19). However,
as explained above, the RHS must be set equal to zero to obtain
the irrep. $\D^-$. The doublet relation in eq. (6.19) should then be
replaced by
\EQ
Q_B||\psi'_1\ra\ra=0.
\EN
Thus $||\psi'_1\ra\ra$ is a singlet in $\D^-$. The ghost excited
state $c_{-1}||\frac{9}{8},\frac{9}{8}\ra\ra$ is then a singlet
in ${\tilde\D}^+$ since its would-be BRST partner is in the different
rep. $\D^-$.

To summarize, in the above example
there is no physical state in the vector
space ${\cal F}^{\pm}_j$. All the states fall into a BRST quartet
rep. and decouple from the physical subspace. In the
PF irreps., however, there are three physical
discrete states $b_{-1}||\frac{9}{8},\frac{9}{8}\ra\ra$,
$||\psi'_1\ra\ra$ and $c_{-1}||\frac{9}{8},\frac{9}{8}\ra\ra$
with ghost number $N_{FP}=0,1$ and 2, respectively.
This conclusion agrees with the general results obtained in sect. 5.

In this way, the physical spectrum may be isomorphic to that of
$c=1$ gravity if we take the vector space ${\cal F}^{\pm}_j$
as a starting point of the string theory on the black hole background.
On the other hand, if we start from the irreps. of
the $SL(2,R)/U(1)$ coset, the spectrum does contain additional infinite
number of discrete states, which we have proved in this paper.

We hope that our discovery of the new physical states provides
some clues to uncover the consistent picture of the quantum
theory of gravity.

\vs{5}
\noindent
{\it Acknowledgements}

We would like to thank M. Kato and Y. Yamada for valuable discussions.
One of the authors (K.I.) would like to
thank Institute for Theoretical Physics at Santa Barbara
and the organizers of the workshop on ``Nonperturbative
String Theory" for their kind hospitality.
This research was supported in part by the National Science Foundation
under Grant No. PHY89-04035.

\newpage
\appendix

{\Large\bf Appendix}

\sect{An alternative formulation of $SL(2,R)/U(1)$}
\indent
In this appendix, we discuss an alternative formalism to realize
reps. of the coset $SL(2,R)/U(1)$.
This is useful for the following reason. The states in PF modules
as defined by $J^3_n=0$, $n\ge 1$
have the same momenta (apart from the factor $i$)
for $\phi^M$ and $\phi^3$ but the operators $S^\pm$ carry only
$\phi^M$ momentum. One may wonder that the mapping of $S^\pm$ is not
closed in the PF modules. This is not a problem because two modules
with different $\phi^3$ momenta give the equivalent reps.
of the same PF module.  We can take the modules with vanishing
$\phi^3$ momentum as the representative of PF
modules in which the $S^\pm$ mapping is closed.
Still it may be better to have a formulation which avoids this
complication. This is what we describe below.

The idea is the following. We introduce an additional free field $X$
as well as two ``fermionic ghosts" $\eta, \xi$, and impose the
constraint that they make a BRST quartet and all decouple from the
theory. Let us define the $U(1)$ BRST charge by
\EQ
Q^{U(1)} = \oint\dz \xi \pa (\phi^3 - iX).
\EN
By imposing the constraint
\EQ
Q^{U(1)}|\psi\ra=0,
\EN
on the reps. $|\psi\ra$ of the current algebra $SL(2,R)$,
we get those of $SL(2,R)/U(1)$.

We define
\bea
{\tilde \psi^\pm} &=& \frac{1}{\sqrt{K}} J^\pm \exp
 \left( \mp i \sqrt{\frac{2}{K}}X \right),  \nonumber \\
 &=& \left(\frac{i}{\sqrt{2}}\pa\phi^M \mp
 \sqrt{\frac{K-2}{2K}}\pa\phi^L\right) \exp
\left( \pm \sqrt{\frac{2}{K}}(i\phi^M+\phi^3 -iX ) \right),\nonumber\\
{\tilde S^\pm} &=& \exp\left( \sqrt{\frac{K-2}{2}} \phi^L \pm
 \sqrt{\frac{K}{2}}( i\phi^M + \phi^3 -iX) \right).
\ena
It is easy to see that
\EQ
[Q^{U(1)},{\tilde \psi^\pm}(z)]=0.
\EN

Due to the constraint (A.2), these operators are equivalent to the
corresponding ones defined in the text. This can be seen by computing
the OPE among the operators.
\bea
{\tilde \psi^\pm}(z){\tilde \psi^\pm}(w) &\sim& {\cal O}((z-w)^{2/K})
 \nonumber \\
 &\sim& \psi^\pm(z) \psi^\pm(w), \nonumber \\
{\tilde \psi^+}(z){\tilde \psi^-}(w) &\sim&
(z-w)^{-2/K} \left[\frac{1}{(z-w)^2}+\frac{\sqrt{\frac{2}{K}}(\pa \phi^3
 -i\pa X)}{z-w} \right. \nonumber \\
&& + \frac{K-2}{K}\left\{ -\shalf(\pa \phi^M)^2 -\shalf
(\pa\phi^L)^2 +\frac{1}{\sqrt{2(K-2)}}\pa^2\phi^L \right\} \nonumber\\
&&\left. + \frac{1}{K}(\pa\phi^3-i\pa X)^2+\frac{1}{\sqrt{2K}}
(\pa^2\phi^3-i\pa^2 X) \right]  \nonumber \\
 &\sim& \psi^+(z) \psi^-(w) + \left\{ Q^{U(1)},
 (z-w)^{-1-2/K}\sqrt{\frac{2}{K}}\eta(w) \right. \nonumber\\
&& \left. + (z-w)^{-2/K}\left( \frac{1}{K}\eta(w)(\pa\phi^3 - i\pa X)
  + \frac{1}{\sqrt{2K}}\pa\eta(w)\right) \right\} .
\ena
where $\psi^\pm$ are the PF's defined in the text.
These relations show that the operators in (A.3) are PF's up to
$U(1)$ BRST-exact form. We also have
\bea
{\tilde \psi^+}(z){\tilde S^+}(w) &\sim& {\rm regular},  \nonumber \\
{\tilde \psi^+}(z){\tilde S^-}(w) &\sim&
 \pa\left[-\frac{1}{\sqrt{K}(z-w)}\exp \left( \sqrt{\frac{K-2}{2}}
\phi^L-\frac{K-2}{\sqrt{2K}}(i\phi^M+\phi^3-iX\right)\right] \nonumber\\
&-& \frac{1}{\sqrt{2}(z-w)} \nonumber\\
&& \times \left\{ Q^{U(1)},
 \eta \exp \left(\sqrt{\frac{K-2}{2}}\phi^L-\frac{K-2}{\sqrt{2K}}
 (i\phi^M+\phi^3-iX)\right) \right\},
\ena
which shows that ${\tilde S^\pm}$ commute with ${\tilde \psi^\pm}$
up to $U(1)$ BRST-exact form.
An important difference from the formulation in the text is that
the reps. in the $SL(2,R)/U(1)$ are defined without spoiling
the dependence on $\phi^M$ and $\phi^3$, and that ${\tilde S^\pm}$
acts in PF modules without the complication noted above.

\sect{Exact sequences}
\indent
In this appendix, we summarize mathematical tools necessary to the
discussions in the text. This is already well known, but we include it
for completeness. We refer the reader to ref.~\cite{BOT} for details.

Our purpose is to examine the structure of the {\it differential complex}
${\cal C}$ which is a direct sum of the vector space ${\cal C}^q$,
($q$: integer) on which the nilpotent differential operator $d$ acts as
\EQ
\cdots \stackrel{d}{\to} {\cal C}^{q-1} \stackrel{d}{\to}
{\cal C}^q \stackrel{d}{\to} {\cal C}^{q+1} \stackrel{d}{\to}
\cdots .
\EN
The object of our interest is the cohomology of ${\cal C}$:
\EQ
H({\cal C})= \bigoplus_{q \in {\bf Z}} H^q({\cal C}),
\EN
where $H^q({\cal C})$ is defined by
\EQ
H^q({\cal C})=Ker (d)\cap{\cal C}^q/ Im (d)\cap{\cal C}^q.
\EN

A map $f:{\cal A} \to {\cal B}$ between two differential complexes is a
{\it chain map} if it commutes with the differential operator $d$.

A sequence of vector spaces $V_i$
\EQ
\cdots \stackrel{f_{i-2}}{\to} V_{i-1} \stackrel{f_{i-1}}{\to}
V_i \stackrel{f_i}{\to} V_{i+1} \stackrel{f_{i+1}}{\to} \cdots
\EN
is said to be {\it exact} if for all $i$
\EQ
Ker(f_i)=Im(f_{i-1}).
\EN

What is most important to us is the {\it short exact sequence}
\EQ
0 \to {\cal A} \stackrel{f}{\to}
{\cal B} \stackrel{g}{\to} {\cal C} \to 0.
\EN
The first part of this sequence means that the exact map $f$ is
one-to-one since it has only 0 as its kernel. Similarly the latter
part implies that the exact map $g$ is onto. Given exact maps $f$
and $g$, ${\cal A}$ is embedded into the space
${\cal B}$ and the rest of the space ${\cal B}$ is mapped to ${\cal C}$.

Given differential complexes ${\cal A, B, C}$ and chain maps
$f:{\cal A}\to {\cal B},g:{\cal B}\to {\cal C}$ which form a short
exact sequence, we can get the {\it long exact sequence} of cohomology
groups
\EQ
\setlength{\unitlength}{1mm}
\begin{picture}(50,50)
\put(0,45){$H^{q+1}({\cal A}) \stackrel{f^*}{\to} \;\; \cdots$}
\put(5,40){$\nwarrow$}
\put(9,39){\line(1,0){30}}
\put(43,35){\line(-1,1){4}}
\put(20,40){$d^*$}
\put(0,27){$H^q({\cal A}) \;\;\stackrel{f^*}{\to} \;\; H^q({\cal B})
 \;\;\stackrel{g^*}{\to} \;\; H^q({\cal C})$}
\put(5,22){$\nwarrow$}
\put(9,21){\line(1,0){30}}
\put(43,17){\line(-1,1){4}}
\put(20,22){$d^*$}
\put(0,10){$H^{q-1}({\cal A}) \stackrel{f^*}{\to} H^{q-1}({\cal B})
 \stackrel{g^*}{\to} H^{q-1}({\cal C})$}
\put(5,5){$\nwarrow$}
\put(10,3){$ \;\; \cdots$}
\end{picture}
\EN

The map $f^*$ is naturally induced as follows. Suppose $f$ maps
an element $a$ of ${\cal A}$ to $b$ in ${\cal B}$.
If $da=0$, one has $db=df(a)=f(da)=f(0)=0$. Also for $a=d\a$, $b=f(d\a)
=df(\a)$. Hence this gives a well-defined map between the elements of
the cohomology groups. Denote their
representatives by $[a],[b]$, respectively.
The map $f^*$ is then defined as
\EQ
f^*: [a] \to [b].
\EN
The map $g^*$ is similarly induced from $g$.

The map $d^*$ is defined as follows. First let us recall the structure
of the map
\EQ
\begin{array}{ccccccccc}
 & & \uparrow d & & \uparrow d &  & \uparrow d & & \\
0 & \to & {\cal A}^{q+1} & \stackrel{f}{\to} & {\cal B}^{q+1}
& \stackrel{g}{\to} & {\cal C}^{q+1} & \to & 0 \\
 & & \uparrow d & & \uparrow d &  & \uparrow d & & \\
0 & \to & {\cal A}^{q} & \stackrel{f}{\to} & {\cal B}^{q}
& \stackrel{g}{\to} & {\cal C}^{q} & \to & 0 \\
 & & \uparrow d & & \uparrow d &  & \uparrow d & &
\end{array} .
\EN
For an element $c \in {\cal C}^{q}, dc=0$, there exist a $b\in {\cal B}
^{q}$ such that $g(b)=c$ since $g$ is onto. Moreover $db\in Ker(g)$
because $0=dc=dg(b)=g(db)$. From the short exact sequence, it follows
that there is an element $a\in {\cal A}^{q+1}$ such that $db=f(a)$.
Since $0=df(a)=f(da)$, we find $da=0$. In this way, we can make a
map between the closed forms in ${\cal A}^{q+1}$ and ${\cal C}^{q}$.

This map is not one-to-one. However, the naturally induced map $d^*:
[c] \to [a]$
gives a homomorphism. Suppose there are two elements $b_1, b_2
\in {\cal B}^q$ which give the same $c\in {\cal C}^q$: $g(b_1)=g(b_2)
=c$. We have $db_1=f(a_1),  db_2=f(a_2)$. Since $0=g(b_1)-g(b_2)=
g(b_1-b_2)$ and hence $b_1-b_2\in Ker(g)=Im(f)$, we find $f(a_{12})
=b_1-b_2$ for some $a_{12}\in{\cal A}^{q+1}$. These relations lead to
$db_1-db_2=f(da_{12})=f(a_1-a_2)$, from which $a_1-a_2=da_{12}$ follows.
Similarly it can be shown that a cohomologically trivial element is
mapped to a trivial element. This completes the proof that $d^*$ is a
homomorphism.

\vs{20}
{\Large\bf Figure captions}

{\large Fig. 1.} The points related by $S^\pm$ to those with $k=l=$
integers. The directions of the map are shown by arrows.

{\large Fig. 2.} BRST nontrivial states in PF modules.
(a) for $N_{FP}=2$, (b) for $N_{FP}=1$, (c) for $N_{FP}=0$.
Here $\cdot$ stands for ${\tilde\D}^+$, $\times$ for $\D^-$,
$\circ$ for ${\tilde\D}^-$, $\bullet$ for $\D^+$,
{\tri} for ${\cal U}$, $\ominus$ for ${\cal C}$.
$\odot$ implies both ${\tilde\D}^+$ and ${\tilde\D}^-$.

{\large Fig. 3(a).} The current algebra module with the Casimir
$-j(j-1)=-\frac{17}{8}\cdot\frac{9}{8}$ generated over the state
with $m=17/8$ ($\otimes$).
Null states are indicated by \tri, $\circ$ and $\bullet$.
This module is related to that in Fig. 3(b) by the Weyl transformation.
States with the same symbols are transformed into each other.

{\large Fig. 3(b).} The module generated over the state with
${\tilde m}=-1$ ($\otimes$). Null states are indicated by \tri, $\circ$
and $\bullet$. Due to the null indicated by \tri, the module is
separated into $\cal U$ and $\tilde \D^+$.

{\large Fig. 4.}
The number of null states on points studied in sect. 3. In a region
separated by two lines, the line to the right included, the number is
as indicated. At points shown with * the embedding structure
and the number of null states are not completely clear.

{\large Fig. 5.} Possible physical states in irreps. (a) for $N_{FP}=2$,
(b) for $N_{FP}=1$, (c) for $N_{FP}=0$, (d) for $N_{FP}=-1$. Here
$\cdot$ stands for ${\hat{\tilde\D}}^+$, $\times$ for ${\hat\D}^-$,
$\circ$ for ${\hat{\tilde\D}}^-$, $\bullet$ for ${\hat\D}^+$,
{\tri} for ${\hat{\cal U}}$, $\ominus$ for ${\hat{\cal C}}$.
$\odot$ implies both ${\hat{\tilde\D}}^+$ and ${\hat{\tilde\D}}^-$.

{\large Fig. 6.} Examples of physical states.

\newpage

\begin{figure}
\setlength{\unitlength}{.7mm}
\begin{picture}(200,130)(-40,-20)
\put(0,0){\vector(1,0){120}}
\put(0,0){\vector(0,1){115}}
\multiput(10,0)(20,0){6}{\vector(-1,2){10}}
\multiput(20,20)(20,0){6}{\vector(-1,2){10}}
\multiput(10,40)(20,0){6}{\vector(-1,2){10}}
\multiput(20,60)(20,0){6}{\vector(-1,2){10}}
\multiput(10,80)(20,0){6}{\vector(-1,2){10}}
\multiput(0,10)(40,0){3}{\vector(2,-1){20}}
\multiput(20,20)(40,0){3}{\vector(2,-1){20}}
\multiput(0,30)(40,0){3}{\vector(2,-1){20}}
\multiput(20,40)(40,0){3}{\vector(2,-1){20}}
\multiput(0,50)(40,0){3}{\vector(2,-1){20}}
\multiput(20,60)(40,0){3}{\vector(2,-1){20}}
\multiput(0,70)(40,0){3}{\vector(2,-1){20}}
\multiput(20,80)(40,0){3}{\vector(2,-1){20}}
\multiput(0,90)(40,0){3}{\vector(2,-1){20}}
\multiput(20,100)(40,0){3}{\vector(2,-1){20}}
\multiput(10,0)(20,0){6}{\circle*{2}}
\multiput(20,0)(40,0){3}{\circle*{2}}
\multiput(0,10)(40,0){4}{\circle*{2}}
\multiput(0,20)(20,0){7}{\circle*{2}}
\multiput(0,30)(40,0){4}{\circle*{2}}
\multiput(10,40)(20,0){6}{\circle*{2}}
\multiput(20,40)(40,0){3}{\circle*{2}}
\multiput(0,50)(40,0){4}{\circle*{2}}
\multiput(0,60)(20,0){7}{\circle*{2}}
\multiput(0,70)(40,0){4}{\circle*{2}}
\multiput(10,80)(20,0){6}{\circle*{2}}
\multiput(20,80)(40,0){3}{\circle*{2}}
\multiput(0,90)(40,0){4}{\circle*{2}}
\multiput(0,100)(20,0){6}{\circle*{2}}
\put(19,-5){1}
\put(39,-5){2}
\put(59,-5){3}
\put(79,-5){4}
\put(99,-5){5}
\put(-5,18){1}
\put(-5,38){2}
\put(-5,58){3}
\put(-5,78){4}
\put(-5,98){5}
\put(20,0){\line(0,1){2}}
\put(40,0){\line(0,1){2}}
\put(60,0){\line(0,1){2}}
\put(80,0){\line(0,1){2}}
\put(100,0){\line(0,1){2}}
\put(0,20){\line(1,0){2}}
\put(0,40){\line(1,0){2}}
\put(0,60){\line(1,0){2}}
\put(0,80){\line(1,0){2}}
\put(0,100){\line(1,0){2}}
\put(125,-3){$k$}
\put(0,125){$l$}
\put(160,40){\vector(2,-1){20}}
\put(160,40){\vector(-1,2){10}}
\put(165,20){$S^+$}
\put(140,45){$S^-$}
\put(160,40){\circle*{2}}
\put(50,-20){{\large Fig. 1}}
\end{picture}
\end{figure}

\begin{figure}
\setlength{\unitlength}{.7mm}
\begin{picture}(130,130)(-60,-20)
\put(0,0){\vector(1,0){120}}
\put(0,0){\vector(0,1){115}}
\multiput(17.5,18)(40,0){3}{$\triangle$}
\multiput(17.5,58)(40,0){3}{$\triangle$}
\multiput(17.5,98)(40,0){3}{$\triangle$}
\multiput(10,40)(20,0){6}{\circle{3}}
\multiput(10,80)(20,0){6}{\circle{3}}
\multiput(40,10)(40,0){2}{\circle*{1}}
\multiput(40,30)(40,0){2}{\circle*{1}}
\multiput(40,50)(40,0){2}{\circle*{1}}
\multiput(40,70)(40,0){2}{\circle*{1}}
\multiput(40,90)(40,0){2}{\circle*{1}}
\multiput(18,38)(40,0){3}{$\times$}
\multiput(18,78)(40,0){3}{$\times$}
\multiput(40,20)(40,0){2}{\circle*{3}}
\multiput(40,60)(40,0){2}{\circle*{3}}
\multiput(40,100)(40,0){2}{\circle*{3}}
\multiput(38,38)(40,0){2}{$\ominus$}
\multiput(38,78)(40,0){2}{$\ominus$}
\put(19,-5){1}
\put(39,-5){2}
\put(59,-5){3}
\put(79,-5){4}
\put(99,-5){5}
\put(-5,18){1}
\put(-5,38){2}
\put(-5,58){3}
\put(-5,78){4}
\put(-5,98){5}
\put(20,0){\line(0,1){2}}
\put(40,0){\line(0,1){2}}
\put(60,0){\line(0,1){2}}
\put(80,0){\line(0,1){2}}
\put(100,0){\line(0,1){2}}
\put(0,20){\line(1,0){2}}
\put(0,40){\line(1,0){2}}
\put(0,60){\line(1,0){2}}
\put(0,80){\line(1,0){2}}
\put(0,100){\line(1,0){2}}
\put(125,0){$k$}
\put(0,125){$l$}
\put(50,-20){\large Fig. 2(a)}
\end{picture}
\end{figure}

\begin{figure}
\setlength{\unitlength}{.7mm}
\begin{picture}(130,130)(-60,-20)
\put(0,0){\vector(1,0){120}}
\put(0,0){\vector(0,1){115}}
\multiput(20,20)(20,0){5}{\circle{3}}
\multiput(20,20)(40,0){3}{\circle*{1}}
\multiput(20,40)(40,0){3}{\circle*{1}}
\multiput(20,60)(20,0){5}{\circle{3}}
\multiput(20,60)(40,0){3}{\circle*{1}}
\multiput(20,80)(40,0){3}{\circle*{1}}
\multiput(20,100)(20,0){5}{\circle{3}}
\multiput(20,100)(40,0){3}{\circle*{1}}
\multiput(38,8)(40,0){2}{$\times$}
\multiput(38,28)(40,0){2}{$\times$}
\multiput(38,48)(40,0){2}{$\times$}
\multiput(38,68)(40,0){2}{$\times$}
\multiput(38,88)(40,0){2}{$\times$}
\multiput(10,40)(20,0){6}{\circle*{3}}
\multiput(10,80)(20,0){6}{\circle*{3}}
\multiput(38,38)(40,0){2}{$\ominus$}
\multiput(38,78)(40,0){2}{$\ominus$}
\multiput(0,20)(0,20){5}{\circle{3}}
\multiput(-2,9)(0,20){5}{$\times$}
\multiput(10,0)(20,0){6}{\circle*{3}}
\multiput(20,0)(20,0){5}{\circle*{1}}
\put(19,-10){1}
\put(39,-10){2}
\put(59,-10){3}
\put(79,-10){4}
\put(99,-10){5}
\put(-10,18){1}
\put(-10,38){2}
\put(-10,58){3}
\put(-10,78){4}
\put(-10,98){5}
\put(20,0){\line(0,1){2}}
\put(40,0){\line(0,1){2}}
\put(60,0){\line(0,1){2}}
\put(80,0){\line(0,1){2}}
\put(100,0){\line(0,1){2}}
\put(0,20){\line(1,0){2}}
\put(0,40){\line(1,0){2}}
\put(0,60){\line(1,0){2}}
\put(0,80){\line(1,0){2}}
\put(0,100){\line(1,0){2}}
\put(125,0){$k$}
\put(0,125){$l$}
\put(50,-20){\large Fig. 2(b)}
\end{picture}
\end{figure}

\begin{figure}
\setlength{\unitlength}{.7mm}
\begin{picture}(130,130)(-60,-20)
\put(0,0){\vector(1,0){120}}
\put(0,0){\vector(0,1){115}}
\multiput(17.5,18)(40,0){3}{$\triangle$}
\multiput(17.5,58)(40,0){3}{$\triangle$}
\multiput(17.5,98)(40,0){3}{$\triangle$}
\put(19,-5){1}
\put(39,-5){2}
\put(59,-5){3}
\put(79,-5){4}
\put(99,-5){5}
\put(-5,18){1}
\put(-5,38){2}
\put(-5,58){3}
\put(-5,78){4}
\put(-5,98){5}
\put(20,0){\line(0,1){2}}
\put(40,0){\line(0,1){2}}
\put(60,0){\line(0,1){2}}
\put(80,0){\line(0,1){2}}
\put(100,0){\line(0,1){2}}
\put(0,20){\line(1,0){2}}
\put(0,40){\line(1,0){2}}
\put(0,60){\line(1,0){2}}
\put(0,80){\line(1,0){2}}
\put(0,100){\line(1,0){2}}
\put(125,0){$k$}
\put(0,125){$l$}
\put(50,-20){\large Fig. 2(c)}
\end{picture}
\end{figure}

\begin{figure}
\setlength{\unitlength}{.8mm}
\begin{picture}(120,120)(-30,-35)
\put(60,0){\line(-1,1){30}}
\put(60,0){\line(1,0){50}}
\put(58,-1){$\otimes$}
\put(115,-7){$m$}
\put(45,-20){\large Fig. 3(a)}
\put(50,60){\line(-1,1){20}}
\put(50,60){\line(1,0){40}}
\put(40,70){\line(1,0){50}}
\put(50,60){\circle{3}}
\put(40,70){\circle*{3}}
\put(100,29){level 3}
\put(100,59){level 9}
\put(100,69){level 10}
\multiput(30,0)(0,5){5}{\line(0,1){2}}
\put(30,25){\vector(0,1){3}}
\put(40,0){\vector(0,1){3}}
\put(50,0){\vector(0,1){3}}
\put(30,0){\line(0,1){2}}
\put(40,0){\line(0,1){2}}
\put(50,0){\line(0,1){2}}
\put(28,-9){$-\frac{7}{8}$}
\put(39,-9){$\frac{1}{8}$}
\put(49,-9){$\frac{9}{8}$}
\put(58,-9){$\frac{17}{8}$}
\put(28,29){\Large\tri}
\thicklines
\put(30,30){\line(-1,1){40}}
\put(30,30){\line(1,0){60}}
\end{picture}
\end{figure}

\begin{figure}
\setlength{\unitlength}{.8mm}
\begin{picture}(120,120)(-30,-35)
\put(30,0){\line(-1,1){30}}
\put(30,0){\line(1,0){20}}
\put(50,0){\line(1,1){30}}
\put(40,50){\line(-1,1){20}}
\put(40,50){\line(1,0){30}}
\put(40,50){\circle{3}}
\put(50,50){\circle*{3}}
\put(80,49){level 8}
\put(50,-20){\large Fig. 3(b)}
\put(40,0){\line(0,1){2}}
\thicklines
\put(60,0){\line(-1,1){60}}
\put(60,0){\line(1,0){40}}
\put(28,-8){-1}
\put(38,-8){0}
\put(48,-8){1}
\put(58,-8){2}
\put(105,-8){$\tilde m$}
\put(28,-1){$\otimes$}
\put(58,-1){\Large\tri}
\end{picture}
\end{figure}

\begin{figure}
\setlength{\unitlength}{.7mm}
\begin{picture}(130,130)(-60,-20)
\put(0,0){\vector(1,0){120}}
\put(0,0){\vector(0,1){115}}
\multiput(40,10)(40,0){2}{\circle*{2}}
\multiput(20,20)(20,0){5}{\circle*{2}}
\multiput(40,30)(40,0){2}{\circle*{2}}
\multiput(10,40)(10,0){3}{\circle*{2}}
\multiput(50,40)(10,0){3}{\circle*{2}}
\multiput(90,40)(10,0){3}{\circle*{2}}
\multiput(40,50)(40,0){2}{\circle*{2}}
\multiput(20,60)(20,0){5}{\circle*{2}}
\multiput(40,70)(40,0){2}{\circle*{2}}
\multiput(10,80)(10,0){3}{\circle*{2}}
\multiput(50,80)(10,0){3}{\circle*{2}}
\multiput(90,80)(10,0){3}{\circle*{2}}
\multiput(40,90)(40,0){2}{\circle*{2}}
\multiput(20,100)(20,0){5}{\circle*{2}}
\multiput(40,110)(40,0){2}{\circle*{2}}
\multiput(38,37)(40,0){2}{*}
\multiput(38,77)(40,0){2}{*}
\put(19,-5){1}
\put(39,-5){2}
\put(59,-5){3}
\put(79,-5){4}
\put(99,-5){5}
\put(-5,18){1}
\put(-5,38){2}
\put(-5,58){3}
\put(-5,78){4}
\put(-5,98){5}
\multiput(5,35)(40,0){3}{\line(1,-1){30}}
\multiput(5,75)(40,0){3}{\line(1,-1){30}}
\multiput(5,115)(40,0){3}{\line(1,-1){30}}
\put(5,10){$s=0$}
\put(22,27){$s=1$}
\put(45,50){$s=2$}
\put(62,67){$s=3$}
\put(85,90){$s=4$}
\put(20,0){\line(0,1){2}}
\put(40,0){\line(0,1){2}}
\put(60,0){\line(0,1){2}}
\put(80,0){\line(0,1){2}}
\put(100,0){\line(0,1){2}}
\put(0,20){\line(1,0){2}}
\put(0,40){\line(1,0){2}}
\put(0,60){\line(1,0){2}}
\put(0,80){\line(1,0){2}}
\put(0,100){\line(1,0){2}}
\put(125,0){$k$}
\put(0,125){$l$}
\put(50,-20){\large Fig. 4}
\end{picture}
\end{figure}

\begin{figure}
\setlength{\unitlength}{.7mm}
\begin{picture}(130,130)(-60,-20)
\put(0,0){\vector(1,0){120}}
\put(0,0){\vector(0,1){115}}
\multiput(17.5,18)(40,0){3}{$\triangle$}
\multiput(17.5,58)(40,0){3}{$\triangle$}
\multiput(17.5,98)(40,0){3}{$\triangle$}
\multiput(10,40)(20,0){6}{\circle{3}}
\multiput(10,80)(20,0){6}{\circle{3}}
\multiput(40,10)(40,0){2}{\circle*{1}}
\multiput(40,30)(40,0){2}{\circle*{1}}
\multiput(40,50)(40,0){2}{\circle*{1}}
\multiput(40,70)(40,0){2}{\circle*{1}}
\multiput(40,90)(40,0){2}{\circle*{1}}
\multiput(18,38)(40,0){3}{$\times$}
\multiput(18,78)(40,0){3}{$\times$}
\multiput(40,20)(40,0){2}{\circle*{3}}
\multiput(40,60)(40,0){2}{\circle*{3}}
\multiput(40,100)(40,0){2}{\circle*{3}}
\multiput(38,38)(40,0){2}{$\ominus$}
\multiput(38,78)(40,0){2}{$\ominus$}
\put(19,-5){1}
\put(39,-5){2}
\put(59,-5){3}
\put(79,-5){4}
\put(99,-5){5}
\put(-5,18){1}
\put(-5,38){2}
\put(-5,58){3}
\put(-5,78){4}
\put(-5,98){5}
\put(20,0){\line(0,1){2}}
\put(40,0){\line(0,1){2}}
\put(60,0){\line(0,1){2}}
\put(80,0){\line(0,1){2}}
\put(100,0){\line(0,1){2}}
\put(0,20){\line(1,0){2}}
\put(0,40){\line(1,0){2}}
\put(0,60){\line(1,0){2}}
\put(0,80){\line(1,0){2}}
\put(0,100){\line(1,0){2}}
\put(125,0){$k$}
\put(0,125){$l$}
\put(50,-20){\large Fig. 5(a)}
\end{picture}
\end{figure}

\begin{figure}
\setlength{\unitlength}{.7mm}
\begin{picture}(130,130)(-60,-20)
\put(0,0){\vector(1,0){120}}
\put(0,0){\vector(0,1){115}}
\multiput(40,20)(40,0){2}{\circle{3}}
\multiput(20,40)(40,0){3}{\circle*{1}}
\multiput(40,60)(40,0){2}{\circle{3}}
\multiput(20,80)(40,0){3}{\circle*{1}}
\multiput(40,100)(40,0){2}{\circle{3}}
\multiput(38,8)(40,0){2}{$\times$}
\multiput(38,28)(40,0){2}{$\times$}
\multiput(38,48)(40,0){2}{$\times$}
\multiput(38,68)(40,0){2}{$\times$}
\multiput(38,88)(40,0){2}{$\times$}
\multiput(10,40)(20,0){6}{\circle*{3}}
\multiput(10,80)(20,0){6}{\circle*{3}}
\multiput(38,38)(40,0){2}{$\ominus$}
\multiput(38,78)(40,0){2}{$\ominus$}
\put(19,-5){1}
\put(39,-5){2}
\put(59,-5){3}
\put(79,-5){4}
\put(99,-5){5}
\put(-5,18){1}
\put(-5,38){2}
\put(-5,58){3}
\put(-5,78){4}
\put(-5,98){5}
\put(20,0){\line(0,1){2}}
\put(40,0){\line(0,1){2}}
\put(60,0){\line(0,1){2}}
\put(80,0){\line(0,1){2}}
\put(100,0){\line(0,1){2}}
\put(0,20){\line(1,0){2}}
\put(0,40){\line(1,0){2}}
\put(0,60){\line(1,0){2}}
\put(0,80){\line(1,0){2}}
\put(0,100){\line(1,0){2}}
\put(125,0){$k$}
\put(0,125){$l$}
\put(50,-20){\large Fig. 5(b)}
\end{picture}
\end{figure}

\begin{figure}
\setlength{\unitlength}{.7mm}
\begin{picture}(130,130)(-60,-20)
\put(0,0){\vector(1,0){120}}
\put(0,0){\vector(0,1){115}}
\multiput(17.5,18)(40,0){3}{$\triangle$}
\multiput(17.5,58)(40,0){3}{$\triangle$}
\multiput(17.5,98)(40,0){3}{$\triangle$}
\multiput(10,40)(20,0){6}{\circle{3}}
\multiput(10,80)(20,0){6}{\circle{3}}
\multiput(40,10)(40,0){2}{\circle*{1}}
\multiput(40,30)(40,0){2}{\circle*{1}}
\multiput(40,50)(40,0){2}{\circle*{1}}
\multiput(40,70)(40,0){2}{\circle*{1}}
\multiput(40,90)(40,0){2}{\circle*{1}}
\multiput(18,38)(40,0){3}{$\times$}
\multiput(18,78)(40,0){3}{$\times$}
\multiput(40,20)(40,0){2}{\circle*{3}}
\multiput(40,60)(40,0){2}{\circle*{3}}
\multiput(40,100)(40,0){2}{\circle*{3}}
\multiput(38,38)(40,0){2}{$\ominus$}
\put(38,78){$\ominus$}
\put(19,-5){1}
\put(39,-5){2}
\put(59,-5){3}
\put(79,-5){4}
\put(99,-5){5}
\put(-5,18){1}
\put(-5,38){2}
\put(-5,58){3}
\put(-5,78){4}
\put(-5,98){5}
\put(20,0){\line(0,1){2}}
\put(40,0){\line(0,1){2}}
\put(60,0){\line(0,1){2}}
\put(80,0){\line(0,1){2}}
\put(100,0){\line(0,1){2}}
\put(0,20){\line(1,0){2}}
\put(0,40){\line(1,0){2}}
\put(0,60){\line(1,0){2}}
\put(0,80){\line(1,0){2}}
\put(0,100){\line(1,0){2}}
\put(125,0){$k$}
\put(0,125){$l$}
\put(50,-20){\large Fig. 5(c)}
\end{picture}
\end{figure}

\begin{figure}
\setlength{\unitlength}{.7mm}
\begin{picture}(130,130)(-60,-20)
\put(0,0){\vector(1,0){120}}
\put(0,0){\vector(0,1){115}}
\multiput(60,40)(40,0){2}{\circle*{1}}
\multiput(40,60)(20,0){4}{\circle{3}}
\multiput(60,80)(40,0){2}{\circle*{1}}
\multiput(40,100)(20,0){4}{\circle{3}}
\multiput(38,48)(40,0){2}{$\times$}
\multiput(38,68)(40,0){2}{$\times$}
\multiput(38,88)(40,0){2}{$\times$}
\multiput(50,40)(20,0){4}{\circle*{3}}
\multiput(50,80)(20,0){4}{\circle*{3}}
\put(19,-5){1}
\put(39,-5){2}
\put(59,-5){3}
\put(79,-5){4}
\put(99,-5){5}
\put(-5,18){1}
\put(-5,38){2}
\put(-5,58){3}
\put(-5,78){4}
\put(-5,98){5}
\put(20,0){\line(0,1){2}}
\put(40,0){\line(0,1){2}}
\put(60,0){\line(0,1){2}}
\put(80,0){\line(0,1){2}}
\put(100,0){\line(0,1){2}}
\put(0,20){\line(1,0){2}}
\put(0,40){\line(1,0){2}}
\put(0,60){\line(1,0){2}}
\put(0,80){\line(1,0){2}}
\put(0,100){\line(1,0){2}}
\put(125,0){$k$}
\put(0,125){$l$}
\put(50,-20){\large Fig. 5(d)}
\end{picture}
\end{figure}

\begin{figure}
\setlength{\unitlength}{1mm}
\begin{picture}(100,100)(-40,-15)
\put(50,0){\line(1,0){50}}
\put(10,40){\line(1,-1){40}}
\put(0,0){\line(1,0){15}}
\put(15,0){\line(1,1){40}}
\put(48,33){\circle{3}}
\put(10,-5){$m=\frac{1}{8}$}
\put(45,-5){$m=\frac{9}{8}$}
\put(0,20){${\cal D}^-$}
\put(70,20){${\tilde{\cal D}}^+$}
\put(30,-15){\large Fig. 6}
\end{picture}
\end{figure}

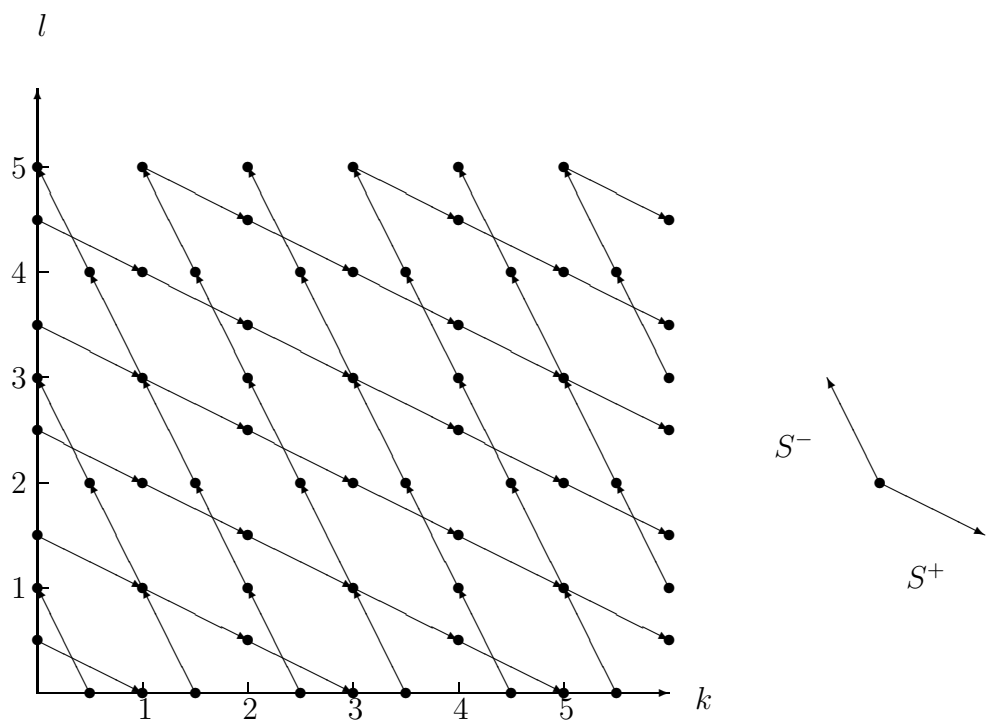
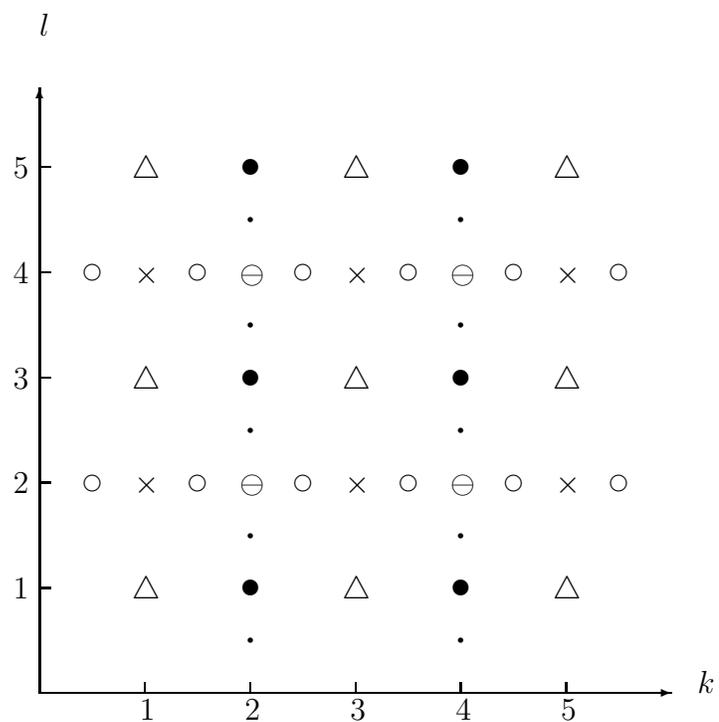
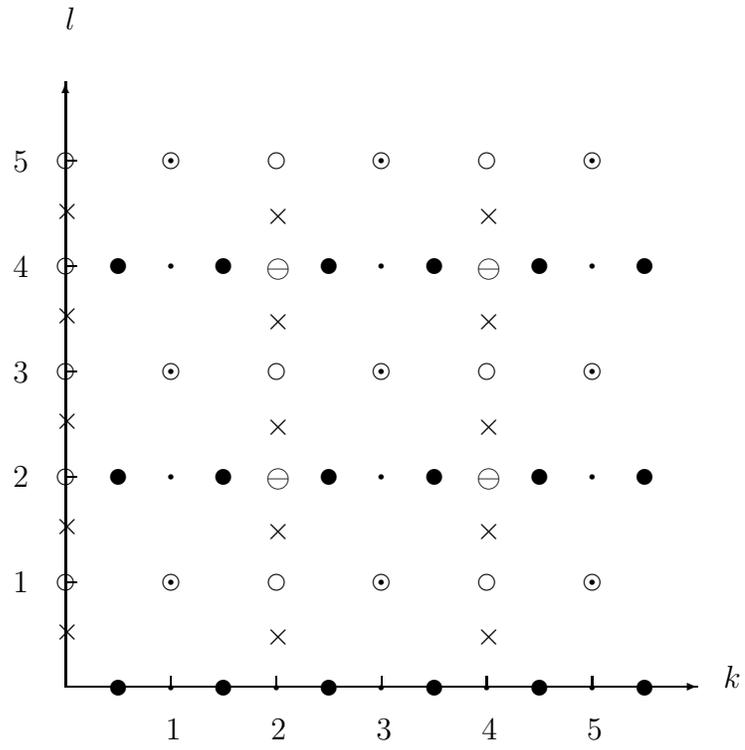
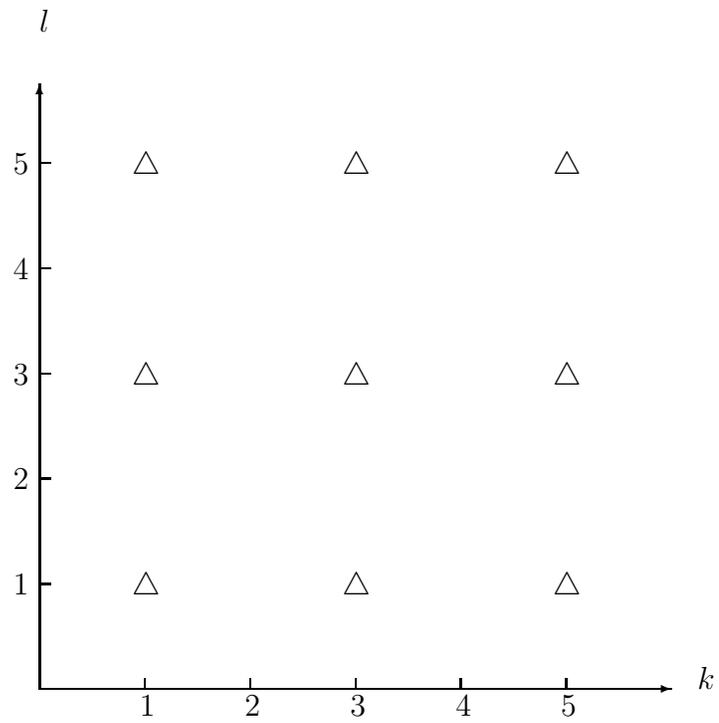
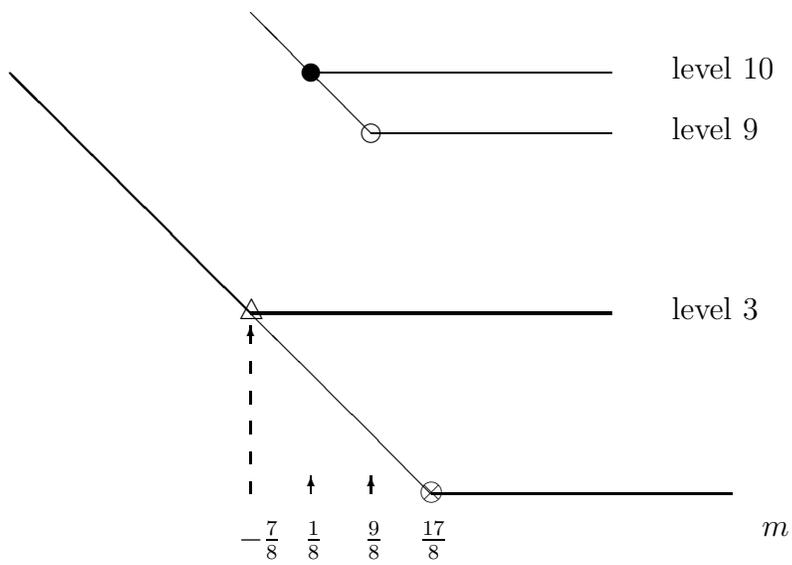
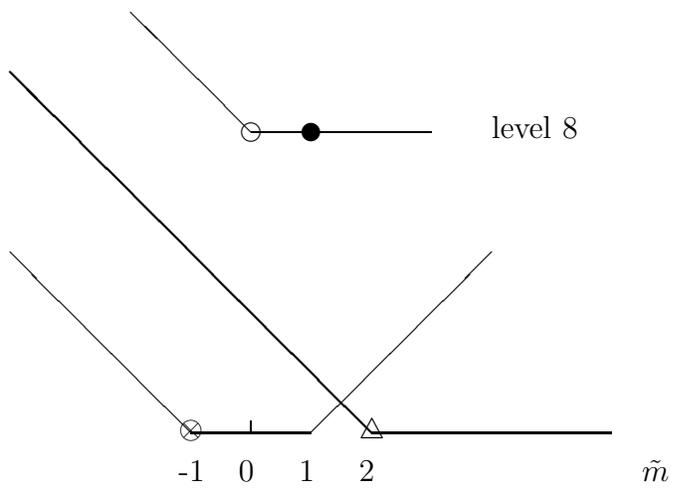
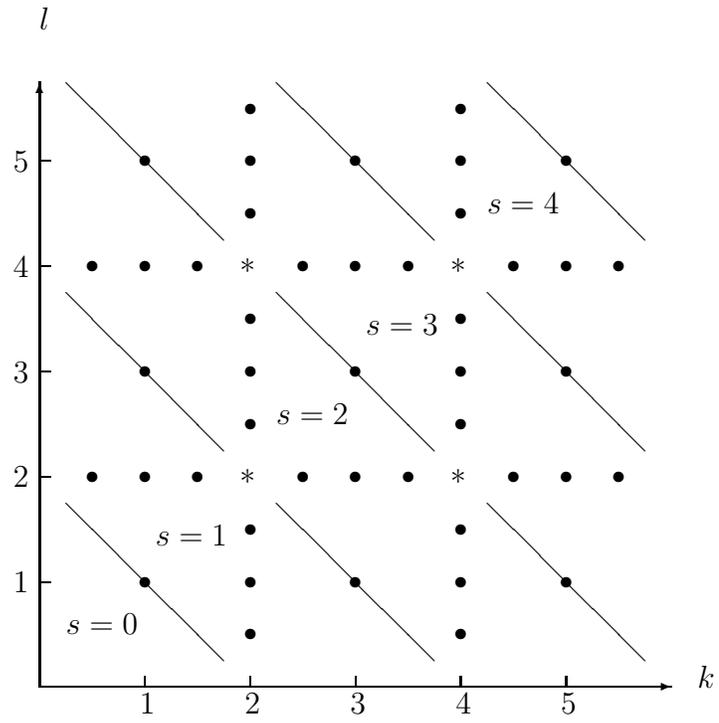
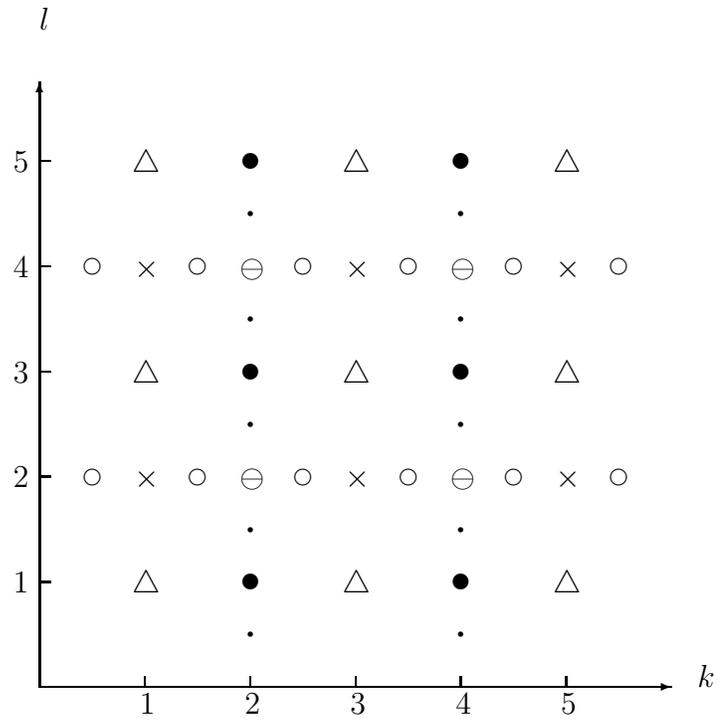
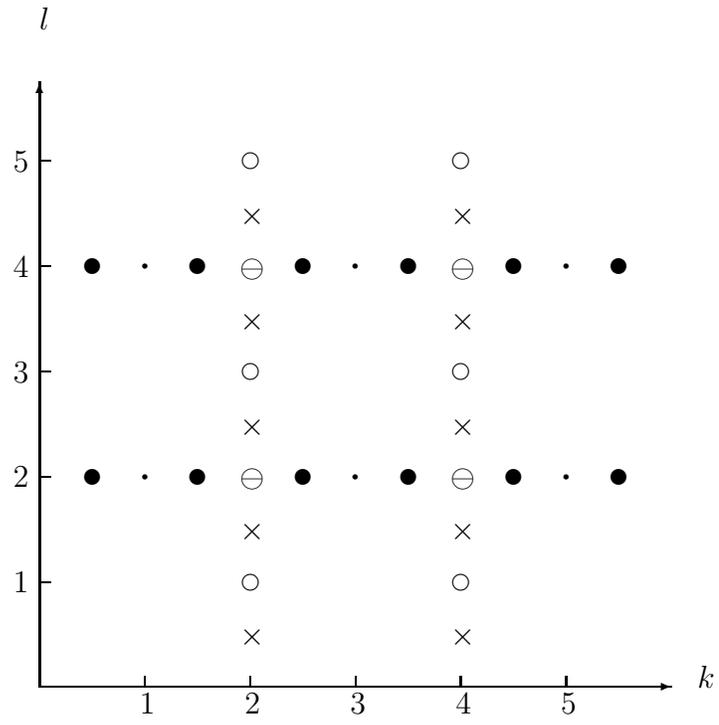
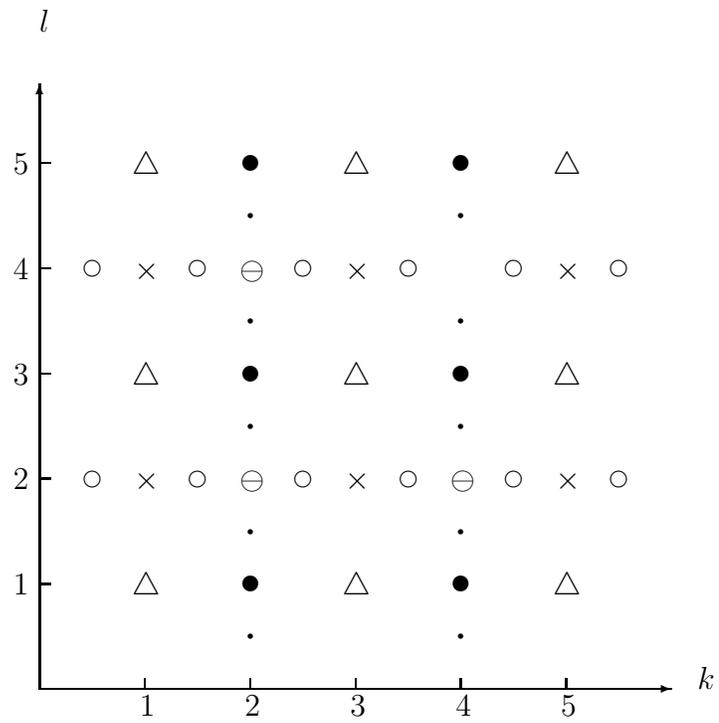
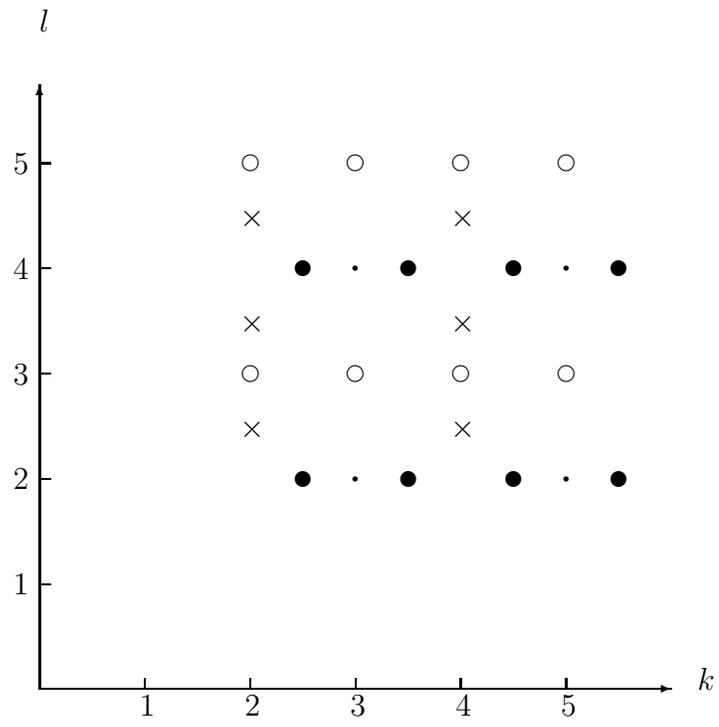
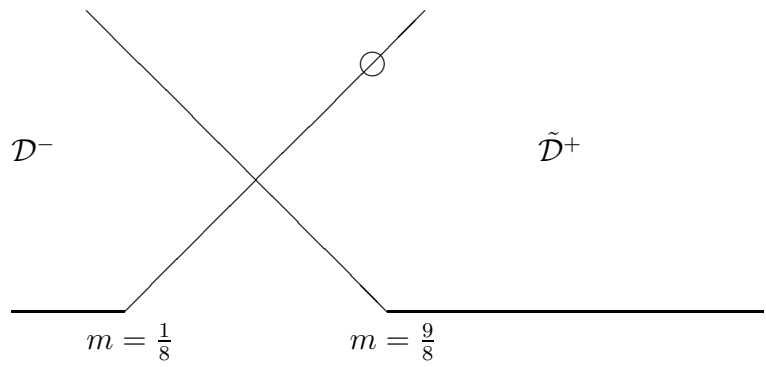
\end{document}